\documentclass[8pt,useAMS,usenatbib]{mn2e}

\newcommand{\etal}{\mbox{et\ al.\ }}

\usepackage{rotating}
\usepackage{url}

\title[A search for thermal X-ray signatures in GRBs I]
{A search for thermal X-ray signatures in Gamma-Ray Bursts I: \emph{Swift} bursts
with optical supernovae}

\author[Starling \etal 2012]
{R.L.C. Starling$^1$\thanks{Royal Society Dorothy Hodgkin Fellow, mailto:rlcs1@le.ac.uk}, K.L. Page$^1$, A. Pe'er$^2$, A.P. Beardmore$^1$ and J.P. Osborne$^1$ \\
$^1$Department of Physics and Astronomy, University of Leicester, University
  Road, Leicester LE1 7RH, UK.\\
$^2$Harvard-Smithsonian Center for Astrophysics, 60 Garden Street, Cambridge, MA 02138, USA.}

\begin{document}
\date{Accepted . Received ; in original form .}

\pagerange{\pageref{firstpage}--\pageref{lastpage}} \pubyear{}

\maketitle

\label{firstpage}


\begin{abstract}
The X-ray spectra of Gamma-Ray Bursts can generally be described by an absorbed power law. 
The landmark discovery of thermal X-ray emission in addition to the power law in the unusual GRB\,060218, followed by a similar discovery in GRB\,100316D, showed that during the first thousand seconds after trigger the soft X-ray spectra can be complex. Both
the origin and prevalence of such spectral components still evade understanding, particularly after the discovery of thermal X-ray emission in the classical GRB 090618. Possibly most importantly, these three objects are all associated with optical supernovae, begging the question of whether the thermal X-ray components could be a result of the GRB-SN connection, possibly in the shock breakout.
We therefore performed a search for blackbody components in the early {\it Swift} X-ray spectra
of 11 GRBs that have or may have associated optical supernovae, accurately recovering the thermal components reported in the literature for GRBs 060218, 090618 and 100316D. We present the discovery of a cooling blackbody in GRB\,101219B/SN2010ma, and in four further GRB-SNe we find an improvement in the fit with a blackbody which we deem possible blackbody candidates due to case-specific caveats. All the possible new blackbody components we report lie at the high end of the luminosity and radius distribution. GRB\,101219B appears to bridge the gap between the low-luminosity and the classical GRB-SNe with thermal emission, and following the blackbody evolution we derive an expansion velocity for this source of order 0.4c. We discuss potential origins for the thermal X-ray emission in our sample, including a cocoon model which we find can accommodate the more extreme physical parameters implied by many of our model fits.
\end{abstract}


\section{Introduction} \label{sec:intro}
The majority of Gamma-Ray Bursts (GRBs), those belonging to the `long' burst class, likely originate in the deaths
of very massive stars. These massive stars end their lives in a Type Ib/c
supernova (SN) explosion, whose signature can be seen at optical wavelengths alongside the GRB emission in a number of
nearby examples with well sampled light curves and/or spectroscopy (e.g. Hjorth \& Bloom 2011).
Supernova signatures have also been claimed in the
high energy emission of the extremely long, subenergetic and rather unusual gamma-ray burst GRB\,060218 in the form of a thermal X-ray component present in the spectrum over the first thousand seconds (Campana et al. 2006) that some attribute to shock breakout of the supernova as it emerges from the star (e.g. Waxman, M\'esz\'aros \& Campana 2007; Nakar \& Sari 2010; Nakar \& Sari 2012), first suggested by Colgate (1974). GRB\,060218 displayed indisputable
spectroscopic evidence for a supernova at optical wavelengths (Pian et al. 2006; Mazzali et al. 2006; Sollerman et al. 2006), while the origin of the thermal X-ray emission remains an open issue. Shock breakout must occur in these systems, but its observable signature is not well known. 

Supernova shock break out may have been observed in a few `ordinary' supernovae at ultraviolet and optical wavelengths (e.g. Schawinski et al. 2008; Gezari et al. 2008,2010; Ofek et al. 2010), but thus far these have been SNe of Type II, and therefore not linked to the GRB systems.  
Shock break out is the interpretation given to a short burst of
X-rays observed from a supernova not associated with a GRB, SN2008D, serendipitously
caught in the act of exploding (Soderberg et al. 2008; Chevalier \& Fransson 2008; Modjaz et al. 2009; Suzuki \& Shigeyama 2010; Balberg \& Loeb 2011; Couch et al. 2011 among others). These serendipitous {\it Swift} data on SN2008D provided the earliest ever observations of a supernova, seen to rise sharply at X-ray energies followed by an exponential decay. In contrast to GRB\,060218, the X-ray spectrum was not thermal but well described by a power law throughout. Shock breakout can lead to either thermal or non-thermal spectra depending on the conditions in the shock and the progenitor star (e.g. Waxman et al. 2007; Nakar \& Sari 2010; Nakar \& Sari 2012), but there are other emission sites from which similar spectra are expected, namely the cocoon surrounding the jet (e.g. Pe'er, M\'esz\'aros \& Rees 2006).
Furthermore, it is unclear whether the energies required to produce the observed thermal X-ray components in GRBs can be obtained through the shock breakout process (e.g. Ghisellini, Ghirlanda \& Tavecchio 2007). These authors instead invoke the central engine to produce the observed
X-ray emission. Recently, the notion of a `failed GRB' has been suggested for low-luminosity GRBs like 060218 (also known as X-ray flashes), in which the GRB jet never breaks out of the star yet the choked jet powers a relativistic shock breakout which emits in X-rays and/or gamma-rays (Bromberg, Nakar \& Piran 2011; Bromberg et al. 2012).

A thermal X-ray component has also been noted in the prompt to afterglow transition phase of GRB\,100316D which displayed similar, unusual high energy properties to GRB\,060218 and is spectroscopically associated with an optical supernova (Starling et al. 2011). The X-ray spectra of these two bursts before about T$_{\rm 0}+$1000\,s cannot be fit
with the absorbed power law or cut-off power law model that describes
the vast majority of GRBs. The addition of a thermal, blackbody
component with a temperature of $\sim$0.1 keV significantly improves the fits. 

So it appeared that very long-duration, sub-energetic, nearby GRBs with associated supernovae harboured a source of thermal X-rays which regular GRBs did not. A subsequently reported detection of a thermal X-ray component in GRB\,090618 (Page et al. 2011) reopened the debate, however, on the origins of such emission. GRB\,090618 is a `typical' GRB in many respects and yet an absorbed power law model was not sufficient to describe its early X-ray spectrum. This GRB has a photometric supernova association (Cano et al. 2011a) and lies relatively nearby, at $z=0.54$, considering the GRB redshift distribution which currently stretches from $0.001<z<9.4$, peaking around $z=2.2$ (e.g. Fynbo et al. 2009; Jakobsson et al. 2012). The large radius and necessarily high luminosity of the blackbody seen in 090618 is a real challenge to many shock breakout models, and in Section \ref{sec:discussion} we will also consider a model for emission from a cocoon surrounding the jet. We note that the early X-ray spectrum of GRB\,101225A may also be better fit with the addition of a thermal component (Campana et al. 2011a; Th\"one et al. 2011), but the redshift is unknown and its classification as a GRB has not been confirmed however. 

Whatever its origin, an additional, likely thermal component is required to
explain the early X-ray emission of at least three GRBs (Campana et al. 2006; Starling et al. 2011; Page et al. 2011). If it is indeed related to
the supernova, and we expect all long GRBs to originate in the deaths of massive stars, we should see
this component in all long GRB which are close enough and the GRB emission is
faint enough for it to be detected. We note, however, the curious cases of long GRBs 060505 and 060614 which possessed all the attributes conducive to a SN search, yet no SN could be found to very deep limits (e.g. Della Valle et al. 2006; Fynbo et al. 2006; Gal-Yam et al. 2006), highlighting the need for a better understanding of the GRB-SN connection.

We attempt here a systematic search for thermal X-ray signatures in a sample of supernova-associated long GRBs observed promptly with {\it Swift's} X-ray Telescope (XRT, Burrows et al. 2005; Gehrels et al. 2004). We describe the sample in Section \ref{sec:sample} and outline the analysis method in Section \ref{sec:method}. Results for each individual source are presented in Section \ref{sec:results} and a supernova-less GRB is discussed in Section \ref{sec:060614}. In Section \ref{sec:discussion} we summarise the overall findings and outline the main caveats. We look at the prompt, afterglow, supernova and host galaxy emission of our sample for any relationship with the presence/absence of thermal X-ray components and we speculate on the origin of GRB-SN thermal X-ray emission. In a second, related paper we extend this to all {\it Swift} GRBs with redshifts, and show simulations intended to reveal the conditions under which the blackbody components we are finding can be reliably recovered (Sparre \& Starling 2012, hereafter Paper II). Our conclusions are summarised in Section \ref{sec:concl}.

\section{Sample selection} \label{sec:sample}
We begin our systematic search for additional, thermal X-ray components by defining a sample of those GRBs with
\begin{enumerate}
\item optical SN signatures, either spectroscopic or photometric,
\item low redshifts, $z\le1$, and
\item sufficient {\it Swift} XRT Windowed Timing mode data up to 1000~s after
  the trigger to create a spectrum with count rate of a few count s$^{-1}$ or more.
\end{enumerate}
The resulting sample includes ten GRBs: 060218, 060729, 070419A, 080319B,
081007, 090618, 100316D, 100418A, 101219B and 120422A. We also include GRB\,091127, for which {\it Swift} observations did not start promptly and the WT data are beyond 1000\,s, since this source shows a very clear supernova signature in the optical (SN2009nz, Cobb et al. 2010; Berger et al. 2011; Vergani et al. 2011), and deserved to be investigated here. It has also been suggested that 091127 could be a higher redshift analogue of the subenergetic bursts such as 980425 and 060218 (Troja et al. 2012).
Unfortunately we could not apply the same argument to GRB\,050525A, associated with SN\,2005nc (Della Valle et al. 2006), because the first {\it Swift} XRT observations of this source from which we could extract a spectrum were both late-time and in PC mode due to the faintness of the source at that time. For completeness we note two further {\it Swift} GRBs with SN associations for which prompt XRT spectra were not collected: 050824 (Sollerman et al. 2007) and 111211A (de Ugarte Postigo, Thoene \& Gorosabel 2012). The sample and their redshifts are listed in Table \ref{tab:knownproperties}.

\begin{table*}
\caption{Properties of the selected GRB-SN sample. Associated supernovae (P = photometric detection, S = spectroscopic detection to be announced), redshift $z$ and Galactic column density $N_{\rm H,Gal}$ (LAB H\,I Survey, Kalberla et al. 2005). We also present here the intrinsic column density $N_{\rm H,int}$ and fit statistic, after fitting the late-time PC mode spectra from the {\it Swift} XRT GRB Repository using Cash statistics (upper limits are shown in italics).\newline
$^*$Footnote to table Redshift references: 1--Mirabal \& Halpern 2006; Sollerman et al. 2006, 2--Fynbo et al. 2009, 3--Berger et al. 2008, 4--Cenko et al. 2009, 5--Cucchiara et al. 2009; Thoene et al. 2009, 6--Starling et al. 2011, 7--de Ugarte Postigo et al. 2011a, 8--de Ugarte Postigo et al. 2011b; Sparre et al. 2011, 9--Tanvir et al. 2012.\newline Supernova references: A--Pian et al. 2006; Sollerman et al. 2006, B--Cano et al. 2011a, C--Hill et al. 2007, D--Bloom et al. 2009, E--Della Valle et al. 2008; Soderberg, Berger \& Fox 2008, F--Cobb et al. 2010, G--Starling et al. 2011, H--Holland et al. 2010; de Ugarte Postigo et al. in preparation, I--Sparre et al. 2011, J--Wiersema et al. 2012; Melandri et al. 2012; Schulze et al. in preparation.}
\label{tab:knownproperties}
\begin{center}
\begin{tabular}{l l l l l l | l l}
GRB & SN& $N_{\rm H,Gal}$ & $z$ &Refs$^*$ & &$N_{\rm H,int}$& Cstat ($dof$)\\
& & $\times$10$^{22}$ && &&$\times$10$^{22}$&\\ 
& & (cm$^{-2}$)  &&&&(cm$^{-2}$)&\\ \hline\hline
060218&  2006aj&  0.094&    0.0331&1,A &&0.58$\pm$0.07 & 360 (326) \\
060729&   P& 0.045&     0.5428&2,B&&0.15$\pm$0.02& 700 (679)\\
070419A&  P&  0.024&     0.9705&2,C&&{\it 1$^{+4}_{-1}$}& 6.4 (4) \\
080319B&  P&  0.011&     0.9382&2,D&&0.12$\pm$0.06& 493 (537)\\
081007&   2008hw& 0.014&     0.5295&3,E&&0.7$\pm$0.2& 193 (276) \\
090618&  P&  0.058&     0.54&4,B&&0.30$\pm$0.04& 573 (673)\\
091127& 2009nz&0.028&0.49&5,F&&0.11$\pm$0.07& 399 (396)\\
100316D& 2010bh&   0.070&    0.0591&6,G&& {\it 0.6$^{+1.2}_{-0.6}$}& 25 (21)\\
100418A&  S&  0.048&     0.6239&7,H&&{\it 0.2$^{+0.7}_{-0.2}$}& 32 (35) \\
101219B & 2010ma&0.031 & 0.5519 &8,I&&0.16$^{+0.17}_{-0.15}$& 138 (165)\\
120422A & 2012bz&0.037 & 0.28 &9,J&& {\it 0.027$^{+0.207}_{-0.027}$}& 67 (84) \\
\end{tabular} 
\end{center}
\end{table*}
\begin{figure*}
\begin{center}
\includegraphics[width=4cm, angle=-90]{fig1a.ps}
\includegraphics[width=4cm, angle=-90]{fig1b.ps}
\includegraphics[width=4cm, angle=-90]{fig1c.ps}
\includegraphics[width=4cm, angle=-90]{fig1d.ps}
\includegraphics[width=4cm, angle=-90]{fig1e.ps}
\includegraphics[width=4cm, angle=-90]{fig1f.ps}
\includegraphics[width=4cm, angle=-90]{fig1g.ps}
\includegraphics[width=4cm, angle=-90]{fig1h.ps}
\includegraphics[width=4cm, angle=-90]{fig1i.ps}
\includegraphics[width=4cm, angle=-90]{fig1j.ps}
\includegraphics[width=4cm, angle=-90]{fig1k.ps}
\caption{Swift XRT observed count rate light curves of the GRB-SN sample, all shown on the same scale and adapted from the XRT GRB Repository. When the WT data were time-sliced, we have indicated the dividing times with grey vertical lines. A colour version appears on-line: in cyan are WT settling mode data, blue are WT mode data (during which we fit to the spectra) and red are PC mode data.}
\label{lcs}
\end{center}
\end{figure*}
\begin{figure}
\begin{center}
\includegraphics[width=8cm, angle=0]{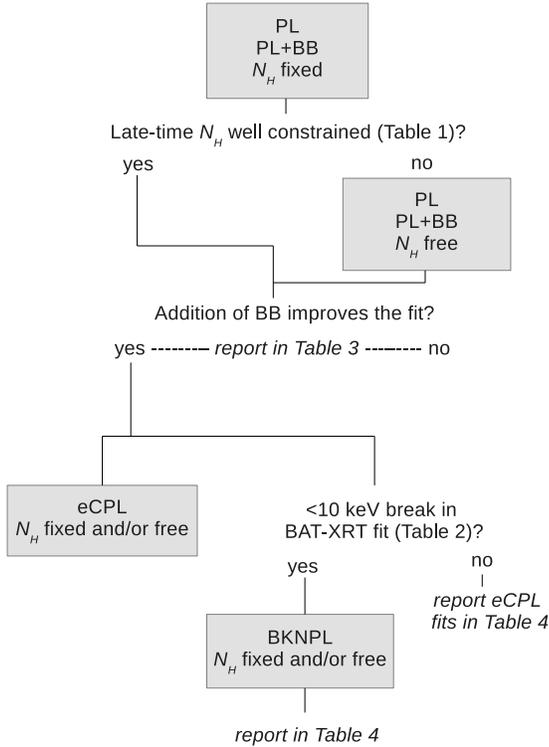}
\caption{Flow chart summarising the sequence of model fits applied to the GRB-SN WT mode spectra.}
\label{flowchart}
\end{center}
\end{figure}

\section{Observations and Method} \label{sec:method}
We obtained the {\it Swift} XRT reduced spectra for all of our GRBs from the UK {\it Swift} Science
Data Centre XRT GRB Repository\footnote{\url{www.swift.ac.uk/xrt_live_cat}}. As a first step, we chose to use the complete time-averaged WT mode spectrum for all sources (excepting GRB\,090618, whose WT data we truncated after the steep decay phase according to Page et al. 2011), ensuring that the data were early-time. Each spectrum was extracted in an identical
fashion, using the {\it Swift} software version 3.8 and following
the method of Evans et al. (2009). The latest calibration
data files (release 20120209) were adopted. The method, including
correction for pile-up, hot columns, vignetting and
exposure maps, is described fully in Evans et al. (2007,2009).

Many GRB time-averaged WT spectra occur during a steep decay phase in the light curve and the time-averaged WT spectrum should be sufficient to determine the need for a thermal component, but some span more complex temporal behaviour.
We examined the XRT GRB Repository$^1$ light curves for the GRB-SN sample to identify any large
flares or other spectral anomalies to be avoided and to choose, where necessary, intervals over which
to extract time-sliced spectra. This provided the time ranges reported in Table 2, column 2.
For those spectra which spanned multiple light curve decay phases (and therefore temporal slopes) we divided the spectrum into time slices where possible (Figure \ref{lcs}). Time slices were chosen to include only one `decay phase' in the temporal behaviour to avoid coadding segments during which the major contributing emission components change.

All WT spectra were grouped such that a minimum of 20 counts occupied each bin. This allowed $\chi^2$ statistics to be used. 

We also obtained the late-time ($>$12h) Photon Counting (PC) mode spectrum for each GRB, and fit these with an absorbed power law model typical of a GRB afterglow, to obtain a
value for the intrinsic X-ray column density, $N_{\rm H,int}$, which we then fixed in our fits to the WT
spectra (listed in Table \ref{tab:knownproperties}). Note that this makes the usual assumption of constant column density throughout our observations (e.g. no detectable ionisation of the circumburst medium). While this is true in general, there are exceptions for which variable columns have been (sometimes tentatively) proposed (e.g. GRB\,011121 - Piro et al. 2005 and GRB\,050904 - Bo\"er et al. 2006; Campana et al. 2007; Gendre et al. 2007; Watson et al. 2007, but see also Butler \& Kocevski 2007).
All of our late-time PC mode fits were carried out on essentially ungrouped spectra (minimum
of 1 count per bin) using Cash statistics (Cash 1979) given the low S/N of a minority of the spectra.

Prompt GRB emission as measured in the hard X-rays/soft gamma-rays is usually well described either by a power law, or by a cut-off power law or Band function when the spectral peak lies within the measured energy range. Early X-ray spectra may fall in a transition period between the prompt and afterglow regimes, leaving the possibility that the spectral peak may enter the XRT band. This is especially relevant for the low-luminosity sources which have lower energy prompt emission. We performed joint fits to both XRT and Burst Alert Telescope (BAT) 15--150 keV spectra where both were available, to more robustly locate the spectral peak energy or other spectral breaks. The BAT data were extracted using the standard {\it Swift} reduction pipelines and the joint spectra were fitted with exponentially cut-off power law models absorbed by a fixed Galactic$+$intrinsic column (Table \ref{tab:knownproperties}), with and without a blackbody. Results of these fits are reported in Table \ref{tab:epk}.

All spectral analysis
was carried out using the X-ray spectral fitting package
Xspec version 12.7.0 (Arnaud 1996).

The sequence of model fits to the WT spectra (both time-averaged and time-sliced) is summarised in Figure \ref{flowchart}. To test for the presence of a thermal component we performed
several fits to the early-time WT spectra of each of
our GRB-SN sample sources, beginning with a power law (PL), then adding in a blackbody (BB) at the source redshift to represent a thermal component and testing for improvement using the F-test (where F-statistic $<0.001$ indicates a significant improvement in fit). Where a BB improved the fit, we continued by removing the BB and replacing the PL with an exponentially cut-off power law (eCPL). If the BAT-XRT joint fits (or literature on the prompt gamma-ray data alone) indicated that a spectral peak may be present in the XRT band (Table \ref{tab:epk}), we also tried a broken power law (BKNPL) model. These models mimic the spectral shape when a break is present, for example the spectral peak energy $E_{\rm pk}$ as often seen in the prompt emission gamma-ray spectrum and shown to be present in the underlying X-ray continua of, among our sample, GRBs 060218 (e.g. Kaneko et al. 2007) and 100316D (Starling et al. 2011). 

Galactic absorption, $N_{\rm H,Gal}$, was kept fixed at the value from the Leiden Argentine Bonn (LAB) H\,I Survey (Kalberla et al. 2005). Additional absorption at the source, $N_{\rm H,int}$, was included with the Xspec {\sc phabs} model using the optically-measured redshifts given in Table \ref{tab:knownproperties}, Solar abundances according to Wilms, Allen \& McCray (2000) and cross-sections derived by Verner et al. (1996). The intrinsic absorption was initially fixed at the late-time-derived value (Table \ref{tab:knownproperties}), and allowed to vary in all cases where the 90\% confidence error on the late-time $N_{\rm H,int}$ measurement was 
$\ge 2 \times$ the value and/or there were fewer than 100 degrees of freedom ($dof$). A grid of fixed $N_{\rm H,int}$ values was used for GRB\,100316D because the late-time PC data were too sparse to obtain an accurate spectral fit.

To further assess the requirement
for the newly reported thermal components we performed monte carlo simulations of an absorbed
PL with fixed $N_{\rm H}$. We created 10000 simulated spectra per GRB using
this model, fitted these with both power law and BB+PL models and deduced the
percentage of these which recover a (non-existent) thermal component by chance. For this number of trials we can at best set a lower limit on the significance of a BB feature of $>4\sigma$, allowing low significances (e.g. $<3\sigma$) to be identified for a modest amount of compute time.
Details of the monte carlo technique can be found in Hurkett et al. (2008) and our results are given in discussions of the individual bursts (Section \ref{sec:results}).

\section{Results} \label{sec:results}
The X-ray fitting results for each source are given hereafter and in Tables \ref{tab:bbfits} and \ref{tab:cutoffpl}, and Figure \ref{ufspec}. We also refer here to BAT-XRT fit results, listed in Table \ref{tab:epk}. GRBs are listed in date order per subsection. 

We find that additional soft X-ray components, described by BBs, are required and favoured in 4 of the GRB-SN sample we investigate, with four further possible BBs. Three of the sources which we find require a BB are confirmations of previous findings: to avoid unnecessary repetition we do not treat these sources in detail but simply show that our findings are consistent. The remaining BB example, GRB\,101219B/SN2010ma, is shown in detail here for the first time. We caution that the thermal emission luminosities and radii we derive here are only valid if the single temperature blackbody is a reasonable approximation and the gas is in thermal equilibrium. 
\begin{table}
\caption{BAT-XRT fits listing the best-fitting model from eCPL or eCPL+BB and
  fitted peak energy, $E_{\rm pk}$. For
  GRB\,100316D we provide estimates for $E_{\rm pk}$ using
  $N_{\rm H,int} = 0, 0.6, 0.91$ and $1.8 \times 10^{22}$ cm$^{-2}$ in that order. }
\label{tab:epk}
\begin{center}
\begin{tabular}{l l l l l}
GRB &BAT-XRT&model&$\chi^2$/$dof$&$E_{\rm pk}$ \\ 
 &overlap (s)&&& (keV)\\ \hline \hline
060218&160-297&eCPL+BB&289/296& 37$^{+35}_{-13}$\\
060729&130-347&eCPL&268/233& unconstrained \\
060729&130-160&eCPL&152/140& 1.1$^{+0.7}_{-0.4}$ \\
060729&195-347&\multicolumn{3}{l}{too weak to fit}\\
070419A&119-145&eCPL&95/113&unconstrained \\
080319B&123-182&eCPL&267/270&7$^{+8}_{-4}$ \\
081007&no overlap&&& \\
090618&125-233&eCPL+BB&208/195 &7$^{+11}_{-4}$\\
091127&no overlap&&& \\
100316D&144-240&eCPL+BB&234/183&21$\pm$3\\
100316D&144-240&eCPL+BB&176/183&19$^{+9}_{-5}$\\
100316D&144-240&eCPL+BB&173/183&19$^{+13}_{-7}$\\
100316D&144-240&eCPL+BB&189/183&21$^{+27}_{-11}$\\
100418A&no overlap&&&\\
101219B&153-540&\multicolumn{3}{l}{too weak to fit}\\
120422A&no overlap&&&\\ 
\end{tabular} 
\end{center}
\end{table}

\subsection{GRB-SNe with thermal X-ray emission}
{\bf GRB\,060218}\\
We recover the thermal X-ray component reported in the literature for GRB\,060218 (e.g. Campana et al. 2006) with high confidence (F-test statistic =  10$^{-105}$). The BB-like emission contributes 13\% of the X-ray flux in the time-averaged WT spectrum. An eCPL cannot replace the addition of a thermal component, and in fact a break in the underlying power law in addition to the BB is required. This confirms previous work, where it is also seen in time-sliced analyses (e.g. Campana et al. 2006; Kaneko et al. 2007). The absorbed BKNPL+BB spectrum we measure has properties consistent with those published elsewhere, validating the method used here. The 0.3--10 keV luminosity we derive for this BB of $\sim 10^{45}$ erg s$^{-1}$ is, together with that of GRB\,100316D, the lowest in the sample, lying approximately 2 orders of magnitude below all the other BB luminosities (see Tables \ref{tab:bbfits} and \ref{tab:bbfinallist}, but note the differing time ranges over which this is calculated).\\ 

\noindent
{\bf GRB\,090618}\\ 
Our PL and PL+BB fits for GRB\,090618 are consistent with the findings of Page et al. (2011) showing an evolving BB component, with a restframe starting temperature of 0.9\,keV cooling to 0.5\,keV at later times, and an underlying spectral softening seen via the power law photon index. The BB is significant at the 99.8\% confidence level in the time-averaged spectrum, and at 99.998\% in the post-flare spectrum.
Our BAT-XRT fits indicate a spectral break $\ge 3$ keV in the overlapping times of 125--233\,s since trigger. Page et al. (2011) found that this could be constrained further when adding in {\it Fermi} GBM data, to lie outside the XRT energy band at 13$\pm$3\,keV. 
 We do not perform further time-sliced analyses since this exists in previously published works. For the purposes of comparison, however, we note that we have fixed $N_{\rm H,int}$ at a slightly different value (using the late-time XRT spectral fit) than did Page et al. (who used the {\it XMM-Newton} spectrum, see also Campana et al. 2011b).\\

\noindent
{\bf GRB\,100316D}\\
This burst is the subject of detailed analyses in Starling et al. (2011) and Olivares et al. (2012) and so only the time-averaged early spectrum is treated here.
The intrinsic X-ray column density cannot be adequately constrained from the late-time spectrum, which returns $N_{\rm H,int}=0.6^{+1.2}_{-0.6} \times 10^{22}$ cm$^{-2}$, so we adopted 4 values for $N_{\rm H,int}$ of 0, 0.6, 0.91 and 1.8 $\times$10$^{22}$ cm$^{-2}$ (from Table \ref{tab:knownproperties} or Starling et al. 2011) to fix in turn. Whichever value we adopt, excepting $N_{\rm H,int} = 0$, a BB is required by the F-test at $\ge 99.6\%$ confidence. We also performed fits with $N_{\rm H,int}$ free to vary and in this case inclusion of the BB improves the fit at the 98.7\% confidence level. A zero intrinsic column density does not provide an acceptable fit with either PL or PL+BB models, hence we conclude that $N_{\rm H,int} = 0$ is excluded.

The spectral peak energy lies outside the XRT band until approximately 240\,s in our BAT-XRT fits, and was shown to remain just outside the XRT band up until at least 737\,s by Starling et al. (2011, their Figure 4 and Table 3). The BAT-XRT fits also all required a BB. All our results hint that this source does indeed have a thermal component in its early evolution, additionally supported by circumstantial evidence from its remarkable similarities to 060218, and evidence for thermal emission moving into the optical bands following the expected cooling (Olivares et al. 2012). Because of the difficulty in assigning a correct $N_{\rm H,int}$ value we do not continue with further more complex fits to this source. The important role that the intrinsic column density plays in determination of any BB component is discussed in detail in Paper II.\\

\noindent
{\bf GRB\,101219B}\\ 
The time-averaged spectrum cannot be modelled with a single absorbed PL, and inclusion of a BB with temperature $kT=0.2$ keV and luminosity $\sim$10$^{47}$ erg s$^{-1}$ improves the fit very significantly. Monte carlo simulations (as described for GRB\,100316D in Starling et al. 2011 and more generally in Hurkett et al. 2008) show that the BB component in this spectrum is $> 4 \sigma$ significant. 
We tested an absorbed eCPL model as well, and this provided no improvement over the PL model. 

Dividing this time interval into three we studied the time-sliced spectra in the same manner. From 160--256\,s since trigger the BB improves the fit at 99.99\% confidence, at 256--320\,s the confidence level is 99.79\% and over 320--544\,s there is improvement at very high confidence, $>$99.99\%, as found in the time-averaged spectrum. The BB temperature decreases with time over the full duration of the WT observations, from 0.30$\pm$0.06 to 0.18$\pm$0.02\,keV, while the luminosity approximately doubles and its contribution to the unabsorbed 0.3--10 keV observer-frame flux goes from 6\% to 25\%. At the same time the power law photon index steepens, seen in a number of GRBs during sharp decays in flux in the prompt to afterglow transition (Zhang et al. 2007). 

All the WT data were taken during an overall steep decay phase in the light curve. This decay continues after the switch to PC mode, and it is therefore interesting to extract the first orbit PC spectrum covering 544--1056\,s and fit the same model. Statistics are poor, and we group the spectrum such that a minimum of 15 counts lie in each bin. If we perform a PL+BB fit we obtain the following BB parameters: $kT=0.10\pm0.02$\,keV, $\Gamma=1.6\pm0.1$, $\chi^2$/$dof$~=~16.2/16 showing a continuation of the cooling measured in earlier data. The BB by this time seems to be fading again, with luminosity similar to that measured in the second time-sliced WT spectrum (but with larger error bar), now contributing 34\% of the flux. 
In Figure \ref{specevol101219b} we show the BB+PL spectral fits to each of the time-sliced spectra described above, and in Figure \ref{bbevol101219b} the evolutionary trends of key spectral parameters over 160--1056\,s are shown. 

\subsection{Possible thermal emission candidates}
{\bf GRB\,060729}\\
A BB+PL is a significantly better fit over a PL for the time-averaged WT spectrum. However, we note a flare in the light curve, which typically produces a hardening in GRB spectra, so we split the spectrum in two segments, pre- and post-flare. The pre-flare spectral fit does not improve with the introduction of a BB component, though it is also not well fit with the absorbed PL model, while the post-flare spectrum does favour such an addition. It is clear that the time-averaged spectrum follows most closely the post-flare spectrum which contains more counts than pre-flare, and we note that the hardness ratio (HR) on the XRT GRB Repository$^1$ shows much spectral softening during the pre-flare spectrum. Monte carlo simulations show that the BB component in the post-flare spectrum is $> 4 \sigma$ significant.
Grupe et al. (2007) performed a detailed analysis of GRB\,060729. They also
found that an additional BB component improves the fits for the later WT
data: although they consider very different time-slices from us, they
state that a BB is required after 150\,s, which is consistent with the
presence of this component in our second time interval of 195--352\,s. Their
range of BB temperatures is also consistent with the mean value determined here.

Over 130--160\,s, during the pre-flare spectrum, we have overlap between the BAT and XRT coverage. Fitting the joint BAT-XRT spectrum reveals a spectral break around 1 keV. Accounting for this using the broken power law model we find the pre-flare XRT spectrum can be well fitted with a break that is consistent with the BAT-XRT joint fit result.
Post-flare we have too little BAT data to perform joint BAT-XRT spectral fits. Fitting the XRT post-flare spectrum with a BKNPL instead of a PL+BB we find that while it remains an improvement over a single PL model, the fit statistic is not as good as the addition of a BB and although $E_{\rm bk}$ moves to lower energies as expected, the spectral slopes are very different from those found pre-flare. 

It is perhaps unusual, given previous examples (although few in number), that a BB is not apparent in the earliest XRT data, but is measureable in subsequent spectra. This is a bright afterglow, which may drown out low luminosity components; the BB is detectable in GRB\,090618 which lies at the same redshift, has a similar shaped light curve and is even brighter in the first 200 seconds, but the BB luminosity we find for 060729 post-flare is only 1/3 of that in 090618. For 060729 we can only conclude that there is significant curvature in the spectrum which may be a hidden BB, but its detection would be complicated by the flaring episode and presence of the spectral peak cascading through the energy band and we are unable to draw firm conclusions. \\

\begin{figure*}
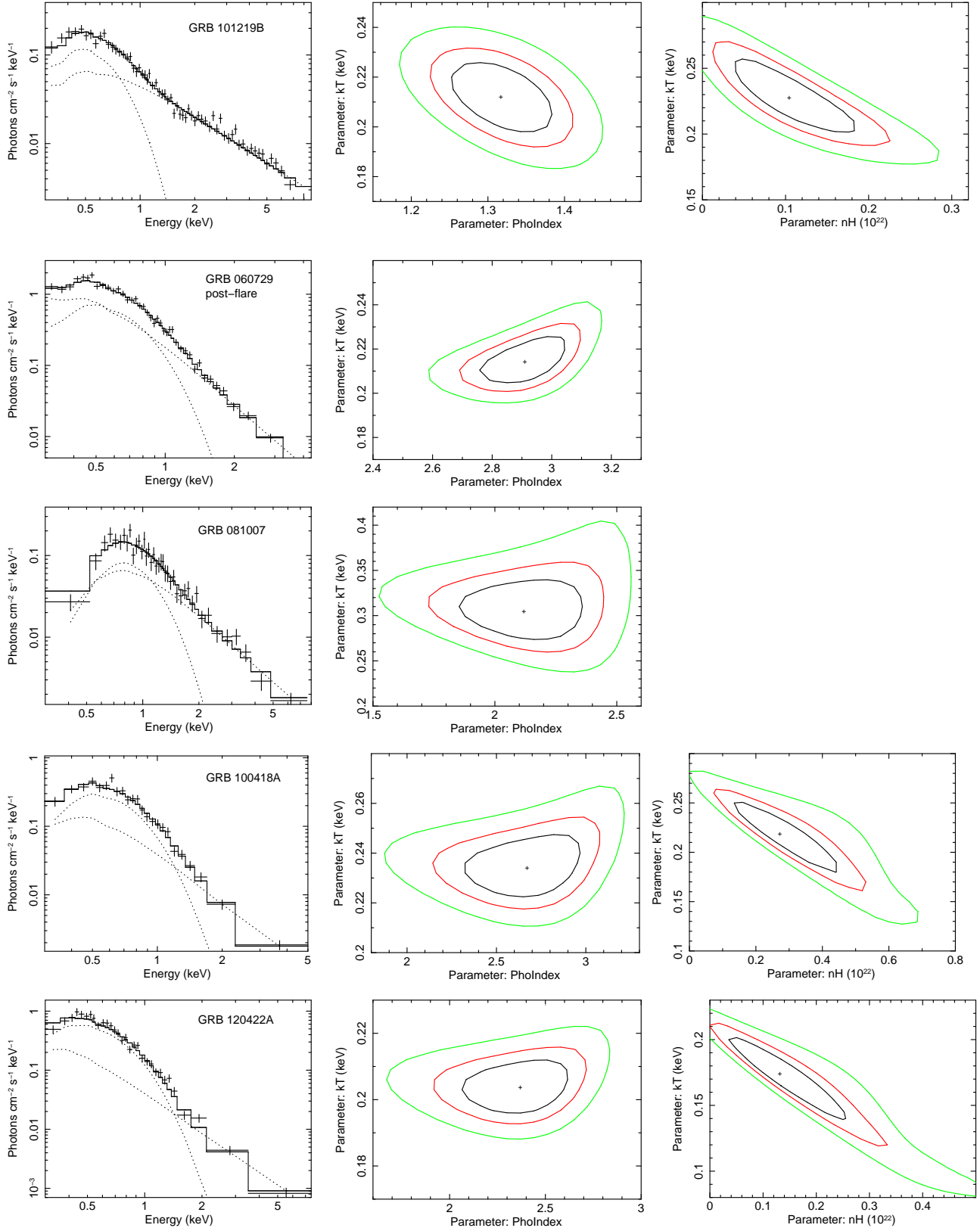

\begin{flushleft}
\caption{On the left hand side we show the unfolded time-averaged WT-mode spectra of GRBs 101219B, 081007, 100418A and 120422A, and the post-flare spectrum of 060729, with the best-fitting absorbed BB+PL model with fixed $N_{\rm H,int}$ overlaid as dotted lines.
To the right of each spectrum we show the 68/90/99\%confidence contour plots for 2 interesting parameters: BB temperature $kT$, against power law photon index $\Gamma$, and $kT$ against intrinsic column density $N_{\rm H,int}$ when that was free to vary.
}
\label{ufspec}
\includegraphics[width=4cm, angle=-90]{fig3a.ps}
\includegraphics[width=4cm, angle=-90]{fig3aa.ps}
\includegraphics[width=4cm, angle=-90]{fig3ab.ps}
\vspace{0.5cm}
\newline
\includegraphics[width=4cm, angle=-90]{fig3b.ps}
\includegraphics[width=4cm, angle=-90]{fig3ba.ps}
\vspace{0.3cm}
\newline
\includegraphics[width=4cm, angle=-90]{fig3c.ps}
\includegraphics[width=4cm, angle=-90]{fig3ca.ps}
\vspace{0.3cm}
\newline
\includegraphics[width=4cm, angle=-90]{fig3d.ps}
\hspace{-0.2cm}
\includegraphics[width=4cm, angle=-90]{fig3da.ps}
\hspace{-0.3cm}
\includegraphics[width=4cm, angle=-90]{fig3db.ps}
\vspace{0.3cm}
\newline
\includegraphics[width=4cm, angle=-90]{fig3e.ps}
\includegraphics[width=4cm, angle=-90]{fig3ea.ps}
\includegraphics[width=4cm, angle=-90]{fig3eb.ps}
\end{flushleft}
\end{figure*}

\begin{figure*}
\begin{center}
\caption{Early X-ray evolution for GRB\,101219B with a PL+BB model fit as listed in Table \ref{tab:bbfits} increasing with time through black--red--green. The data are displayed as $E^2F_{E}$ vs $E$, equivalent to $\nu F_{\nu}$. The blue data show the addition of the PC-mode spectrum, which appears to be a good extrapolation of the earlier evolution.}
\label{specevol101219b}
\includegraphics[width=11cm, angle=-90]{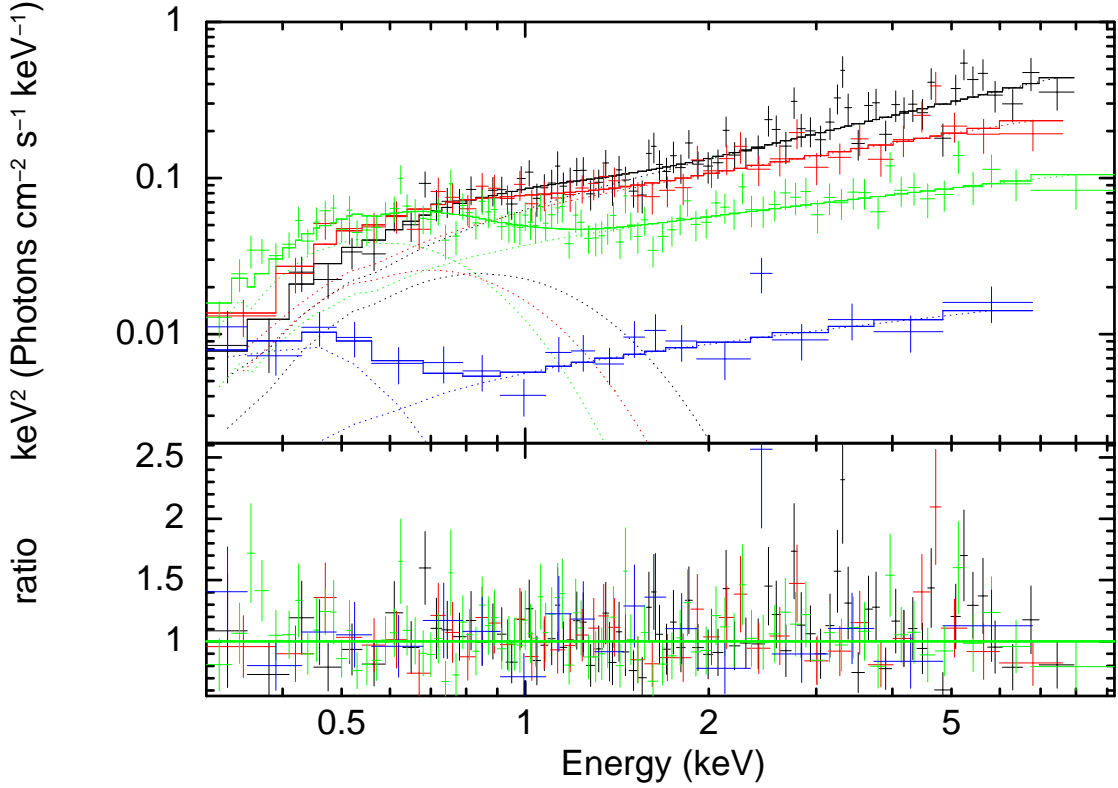}
\end{center}
\end{figure*}

\noindent
{\bf GRB\,081007}\\ 
A model comprising a PL+BB is a better fit than a single PL for this source with significance 99.99\%. In monte carlo simulations of an absorbed power law, we found a chance improvement in the fit with a BB in 6/9990 trials, reducing the significance of the result for 081007 in Table \ref{tab:bbfits} to 99.94\%. Our eCPL fits, with and without a BB, did not improve the fit statistic because the fitted cut-off energy was unconstrained, tending to far beyond the XRT energy band consistent with the single PL fit to prompt {\it Swift} BAT and {\it Fermi} GBM spectra (Markwardt et al. 2008; Bissaldi, McBreen \& Connaughton 2008). The XRT spectrum has relatively few total counts, resulting in substantial errors on the power law slope when a BB is included in the fixed $N_{\rm H,int}$ fit (Figure \ref{ufspec}). We note that $N_{\rm H,int}$ is reasonably large, which may present difficulties in identifying any BB component (see Paper II). While we only show the fixed $N_{\rm H,int}$ fits in Table \ref{tab:bbfits} (as the late-time-derived $N_{\rm H,int}$ is known to within $\sim$30\%), allowing $N_{\rm H,int}$ to vary also resulted in an acceptable fit with a steeper power law photon index and no need for a BB. Intrinsic column density in this case increases, but remains just consistent at the 90\% level with the late-time fit. For this reason, we can only report here a possible, unconfirmed BB component. This is one of the fainter sources in WT mode in our sample.\\

\noindent
{\bf GRB\,100418A}\\ 
The addition of a BB when the column density is fixed resulted in a very significant improvement in the fit statistic and monte carlo simulations show that the BB component in this spectrum is $> 4 \sigma$ significant. However, the uncertainties on $N_{\rm H,int}$ are large and allowing this to be a free parameter in the PL model also provided a significant improvement to the fit while increasing the column density from the fixed value of 2$\times$10$^{21}$ to (6$\pm$1)$\times$ 10$^{21}$ cm$^{-2}$. The addition of a BB at that stage provided another small improvement in the fit, but interestingly the column density then reverts to around its originally fixed value.
An eCPL model with fixed $N_{\rm H,int}$ is not an adequate fit. With $N_{\rm H,int}$ left free (again converging to $\sim$6$\times$10$^{21}$ cm$^{-2}$) this improves, but the cut-off lies way outside the considered energy band meaning we are essentially just fitting a power law as before. This is consistent with the single PL fits reported for prompt BAT data (Ukwatta et al. 2010). Trying instead a BKNPL, we find an adequate fit with $E_{\rm bk} = 0.6\pm0.1$\,keV. It is very difficult, however, to constrain a break at such low energy. If we then allow $N_{\rm H,int}$ to vary it is apparent that $N_{\rm H,int}$ and $\Gamma_1$ are degenerate (Figure \ref{ufspec}) so we do not consider this result further. In conclusion, the PL+BB model with or without $N_{\rm H,int}$ free has the best fit statistic and simple alteration of the PL model with free $N_{\rm H,int}$ parameter values alone does not seem to achieve this but also results in an acceptable fit.\\

\noindent
{\bf GRB\,120422A}\\
The WT spectrum cannot be modelled with a single absorbed PL. Inclusion of a BB significantly improves the fit and monte carlo simulations show that the BB component in this spectrum is $> 4 \sigma$ significant. However, residuals still remain and $\chi^2_{\rm reduced} \sim 1.3$. If the intrinsic absorbing column is allowed to vary, as shown in Table \ref{tab:bbfits} we still prefer the inclusion of a BB both due to improvement in the fit and to alleviate the need for a very steep PL photon index. The fit statistic for the BB+PL model with $N_{\rm H,int}$ free, however, is of the same order as that when $N_{\rm H,int}$ is fixed, and $N_{\rm H,int}$ is poorly constrained. The restframe BB temperature is of order 0.2\,keV with a luminosity of $\sim$10$^{47}$ erg s$^{-1}$, contributing 50--60\% of the unabsorbed 0.3--10\,keV flux. 

We do not have any BAT-XRT overlap or detection in the literature of a spectral peak energy in the prompt phase. An eCPL model does not provide an improvement over the single PL, with the peak energy coverging to a value high above the XRT band. Instead we tried a BKNPL - the fit statistic is acceptable, though not quite as good as the BB+PL fits. It resulted in a break energy of 0.6\,keV and a poorly constrained $\Gamma_1$. The BB+PL model is preferred for this source, but a BKNPL and/or increase in the very low intrinsic column cannot be ruled out. We note that, using a cut-off PL model, Zhang et al. (2012) suggest this source is inconsistent with shock breakout models and suggest instead an engine-driven GRB with steep X-ray decay caused by the curvature effect. 

\subsection{GRB-SNe with no detectable thermal X-ray emission}
{\bf GRB\,070419A}\\
In addition to the time-averaged WT spectrum we created a further time-selected spectrum for GRB\,070419A to avoid early flaring. In both spectra a fixed $N_{\rm H,int}$ PL+BB is a better fit than a single PL with F-test statistic of $\sim$5e-3. The BB temperatures of 0.1--0.2\,keV we measure are similar to those found in GRB\,060218, while the BB luminosity is 1000 times greater. However, the intrinsic column density is not known to any reasonable degree of accuracy due to poor S/N in the late-time spectrum so we also performed a fit with $N_{\rm H,int}$ allowed to vary. In these latter fits we find that an absorbed PL is a perfectly adequate description of the data: the fit statistic improves significantly and the intrinsic column density converges to a lower value than previously fixed at while still remaining consistent with $N_{\rm H,latetime}$ (as expected given its large error bars). 

BAT-XRT fits in the temporal overlap of 119--145\,s (covering some of the time-averaged spectrum but only during the flaring episode) did not constrain any spectral break. An eCPL fit to the XRT spectra after the flare shows that with fixed $N_{\rm H,int}$ we can obtain a very good fit with peak energy $E_{\rm pk}$ around 4--7\,keV (no BB needed) and $\Gamma_1$ steepens a little compared with the PL-only fits, while when $N_{\rm H,int}$ is left free to vary $E_{\rm pk}$ cannot be constrained since there are likely too few photons in the energy bins above a few keV. In either case the eCPL fits are of similar fit statistic to the PL+BB in the fixed $N_{\rm H,int}$ case or the PL in the free $N_{\rm H,int}$ case, thus we conclude that there is no evidence to support the presence of a thermal X-ray component in this source.
This is the most distant GRB-SN in our sample at $z\sim0.97$.\\

\noindent
{\bf GRB\,080319B}\\ 
There are a lot of WT data for this bright burst, therefore we created 3 time-sliced spectra (split by light curve decay segments, Fig \ref{lcs}.) in addition to the time-averaged spectrum and fit all four. We conclude here that no BB is required. However, the single absorbed PL model is a poor fit to the data from the time-averaged and final time-slice spectra, which resemble each other closely since the latest time-slice we have chosen contains a much larger portion of the total WT mode exposure time. We therefore tried fitting all of our 080319B spectra with the eCPL model. This marginally improves the fits overall, but greatly improves the fit to the time-averaged and final time-slice spectra, with the peak energy seeming to cascade downwards in energy from about 7 to 4\,keV. For the final time slice we also tested a BKNPL because the peak energy fell in the mid range of our spectral energy band. This further improved the fit and brought the spectral break to 1--2\,keV. In all these cases we checked the result of letting $N_{\rm H,int}$ vary, and the value did not stray far from its late-time measurement. In summary, the early X-ray data of 080319B can be fit with a set of power laws with a spectral break that moves through the X-ray band to lower energies becoming clearly measurable between 704 and 1744\,s. Our BAT-XRT fits support the presence of a low energy spectral peak, measuring $3 \le E_{\rm pk} \le 15$ keV at 123--182\,s.  

Racusin et al. (2008) present a two-jet model to explain the shape of the X-ray afterglow temporal decay. We note that the portion of the light curve we sample here is dominated by the narrow jet in their model (the wide jet contributes approximately 1/100 of the flux in their Supplementary Figure 7) and does not undergo a temporal break until after our observations. In our search for a thermal component we should therefore not be confused by temporal breaks or interplay between the proposed two jet components according to their model. These authors require the cooling break for the narrow jet forward shock to lie above the X-ray band throughout the time covered by our X-ray observations, ruling out $\nu_c$ as the break in our fits. They also state that the spectral peak of the narrow jet must cross the X-ray band prior to 60\,s when XRT observations began, also ruling out $\nu_m$ as the break suggested in our fits.

Bloom et al. (2009), who proposed the existence of an optical supernova bump in the light curve (and hence this sources' inclusion in this paper), could not fit the early X-ray emission from 080319B within the standard model. Their preferred model while different from that of Racusin et al. also requires that the cooling frequency lies above the X-ray regime for the duration of the observable X-ray afterglow, again ruling out $\nu_c$ as the break suggested from our fitting. However, these authors focussed on the optical, infrared and PC-mode X-ray data and did not attempt to explain the early X-ray data, suggesting it is unrelated to the early optical emission since the two have very different behaviours, or at least there is a spectral break between those bands which moves to higher frequencies with time. We do not see a break moving in the direction of higher frequencies, but if breaks within the X-ray band, thermal components or other spectral curvature were not taken into account then this may explain the difficulty in fitting these data.

We also point out that the redshift of this GRB is the second highest included, at $z\sim0.94$, and that the photometric identification of a supernova in this source is highly uncertain.\\

\noindent
{\bf GRB\,091127}\\ 
The spectrum of GRB\,091127 is adequately fit with an absorbed PL. Given the post-3000\,s observation start, much later than any of the other sample objects, it may be too late to see any detectable thermal component.

\begin{table*}
\caption{Table of results from absorbed PL and absorbed PL+BB fits to the GRB-SN sample. Column 1 also lists time since BAT trigger covered by the spectrum. f denotes a fixed column density. $\frac{F_{\rm BB}}{F_{\rm X}}$ refers to the fraction of the total observer frame 0.3-10\,keV ratio of unabsorbed flux contributed by the BB component. The total luminosity of the BB component is calculated from the BB normalisation and luminosity distance, as given in Page et al. (2011).
F-statistic gives the F-Test probability of a chance improvement in the fit when adding a BB component (or freeing $N_{\rm H,int}$ if given in brackets).}
\label{tab:bbfits}
\begin{center}
\begin{tabular}{l l l l l l l l l }
GRB:time & model& $N_{\rm H,int}$& $kT_{\rm rest}$& $\Gamma$ & $\frac{F_{\rm BB}}{F_{\rm X}}$ & $L_{\rm BB}$& $\frac{\chi^2}{dof}$&Fstat\\ 
 &      & $\times$10$^{22}$ & & & & $\times$10$^{47}$ & & \\ 
~~~~~(s)  &      & (cm$^{-2}$)& (keV)& &(\%) & (erg s$^{-1}$) & & \\ \hline \hline
060218:159-2783&PL& 0.58 f &&       1.820$\pm$0.007 &&& 2826/788 & \\ 
060218:159-2783&PL+BB& 0.58 f &      0.181$\pm$0.004 &       1.63$\pm$0.01 &  13 & 0.024$\pm$0.001 & 1528/786 & 10$^{-105}$\\ 
\hline \hline
060729:128-352 &PL& 0.15 f &&       2.82$\pm$0.03 &&& 272/169 & \\ 
060729:128-352 &PL+BB& 0.15 f &      0.302$^{+0.027}_{-0.022}$ &       2.74$\pm$0.07 &   20 & 16$^{+4}_{-3}$ & 184/167 & 10$^{-15}$\\ 
\hline
060729:128-160 &PL& 0.15 f &&       1.98$\pm$0.06 &&& 230/69 & \\ 
060729:128-160 &PL+BB& 0.15 f &      0.52$^{+0.05}_{-0.04}$ & 1.7$^{+0.2}_{-0.3}$ &    & 
& 230/67 & -\\ 
060729:195-352 &PL& 0.15 f &&       3.25$\pm$0.05 &&& 230/118 & \\ 
060729:195-352 &PL+BB& 0.15 f &      0.21$\pm$0.01 &       2.9$\pm$0.2 &  37 & 22$\pm$5 & 145/116 & 10$^{-12}$\\ 
\hline \hline
070419A:112-304 &PL& 1 f &&        2.3$^{+0.05}_{-0.049}$ &&& 179/166 & \\ 
070419A:112-304 &PL+BB& 1 f &      0.19$\pm$0.08 &       2.19$^{+ 0.08}_{-0.10}$ &   11 & 96$^{+265}_{-50}$ & 168/164 & 6$\times$10$^{-3}$\\ 
070419A:112-304 &PL&      0.81$^{+0.09}_{-0.08}$ &&       2.19$\pm$0.07 &&& 167/165 & (7$\times$10$^{-4}$)\\ 
070419A:112-304 &PL+BB&      0.76$^{+0.22}_{-0.17}$ &      0.4$^{+0.2}_{-0.3}$ &       2.1$^{+0.1}_{-0.2}$ &  6 & 31$\pm$31 & 166/163 &0.61 \\ \hline
070419A:150-304 &PL& 1 f &&       2.42$\pm$0.06 &&& 119/111 & \\ 
070419A:150-304 &PL+BB& 1 f &      0.15$^{+0.06}_{-0.05}$ &       2.31$\pm$0.09  &  16 &177$^{+490}_{-102}$& 108/109 & 5$\times$10$^{-3}$\\ 
070419A:150-304 &PL&      0.75$^{+0.1}_{-0.09}$ &&       2.28$^{+0.09}_{-0.08}$ &&& 105/110 & (2$\times$10$^{-4}$)\\ 
070419A:150-304 &PL+BB&      0.7$^{+0.3}_{-0.2}$ &      0.4$^{+0.6}_{-0.4}$ &       2.22$^{+0.42}_{-0.01}$ &   & & 105/108 & - \\ \hline \hline
080319B:60-1744 &PL& 0.12 f &&       1.67$\pm$0.008 &&& 1213/592 & \\ 
080319B:60-1744 &PL+BB& 0.12 f &      0.93$\pm$0.05 &       1.67$\pm$0.015 &   &  & 1214/590 & -\\ 
\hline
080319B:60-304 &PL& 0.12 f & &      1.71$\pm$0.03 &&& 295/266 & \\ 
080319B:60-304 &PL+BB& 0.12 f &      1.0$\pm$0.4 &       1.72$\pm$0.04 &  4 & 420$^{+316}_{-255}$ & 288/264 & 0.04\\ 
080319B:304-704 &PL& 0.12 f &&       1.64$\pm$0.02 &&& 486/386 & \\ 
080319B:304-704 &PL+BB& 0.12 f &      0.8$\pm$0.1 &       1.62$\pm$0.03 &  &  & 486/384 & -\\ 
080319B:704-1744 &PL& 0.12 f &      & 1.660$\pm$0.008 &&& 1274/572 & \\ 
080319B:704-1744 &PL+BB& 0.12 f &      0.92$\pm$0.04 &       1.66$\pm$0.02 &  &  & 1274/570 & - \\ 
\hline \hline
081007:105-175&PL& 0.7 f &&       2.55$\pm$0.09 &&& 45.6/38 & \\ 
081007:105-175&PL+BB& 0.7 f &      0.31$\pm$0.04 &  2.1$\pm$0.3 &  32 & 6$^{+2}_{-3}$ & 29.7/36 & 4$\times$10$^{-4}$\\ 
\hline \hline
090618:125-250 &PL& 0.3 f &    &   1.45$\pm$0.05 &&& 145/133 & \\ 
090618:125-250 &PL+BB& 0.3 f &      0.9$^{+0.3}_{-0.2}$ &       1.42$\pm$0.07 &  10 & 181$^{+62}_{-53}$ & 132/131 & 2$\times$10$^{-3}$\\ 
\hline
090618:160-250 &PL& 0.3 f &     &  1.9$\pm$0.04 &&& 134/131 & \\ 
090618:160-250 &PL+BB& 0.3 f &      0.51$^{+0.08}_{-0.05}$ &       1.83$\pm$0.06 &  11& 61$^{+15}_{-14}$ & 113/129 & 2$\times$10$^{-5}$\\ \hline \hline
091127:3232-3840 &PL& 0.11 f &&       1.77$\pm$0.04 &&& 158/176 & \\ 
091127:3232-3840 &PL+BB& 0.11 f &      0.6$^{+0.3}_{-0.2}$ &       1.75$\pm$0.06 &   & & 158/174 & -\\ 
\hline \hline
100316D:144-736&PL& 0 f &&       0.42$\pm$0.02 &&& 3628/491 & \\ 
100316D:144-736&PL+BB& 0 f &       0.001$^{+0.037}_{-0.001}$ &      0.42$\pm$0.02 &  $<$1 & - & 3628/489 & -\\ 
100316D:144-736&PL& 0.6 f &&       1.11$\pm$0.02 &&& 826/491 & \\ 
100316D:144-736&PL+BB& 0.6 f &       0.72$\pm$0.02 &      0.80$\pm$0.05 &  20 & 0.046$\pm$0.006 & 590/489 & 10$^{-36}$\\ 
100316D:144-736&PL& 0.91 f &&       1.33$\pm$0.02 &&& 536/491 & \\ 
100316D:144-736&PL+BB& 0.91 f &     0.58$^{+0.12}_{-0.06}$ &       1.24$\pm$0.04 & 5 & 0.012$^{+0.007}_{-0.006}$ & 524/489 & 4$\times$10$^{-3}$\\
100316D:144-736&PL& 1.8 f &&       1.77$\pm$0.03 &&& 832/491 & \\ 
100316D:144-736&PL+BB& 1.8 f &      0.176$\pm$0.009 &        1.50$\pm$0.03 & 33 & 0.16$^{+0.05}_{-0.03}$ & 512/489 & 10$^{-52}$\\ 
100316D:144-736&PL&       1.07$\pm$0.05 &&       1.43$\pm$0.04 &&& 508/490 & \\ 
100316D:144-736&PL+BB&       1.27$^{+0.14}_{-0.10}$ &       0.23$^{+ 0.04}_{-0.03}$ &  1.4$^{+0.03}_{-0.04}$ &  8& 0.026$^{+0.036}_{-0.017}$ & 499/488 & 0.013\\ \hline \hline
100418A:64-160&PL& 0.2 f &&       3.27$\pm$0.08 &&& 94.9/46 & \\ 
100418A:64-160&PL+BB& 0.2 f &      0.23$\pm$0.01 &  2.7$\pm$0.2 &  57& 16$\pm$4 & 34.7/44 & 10$^{-10}$\\ 
100418A:64-160&PL&      0.6$\pm$0.1 &&        4.2$\pm$0.3 &&& 46.0/45 & (10$^{-8}$)\\ 
100418A:64-160&PL+BB&      0.28$^{+0.19}_{-0.16}$ &  0.22$\pm$0.04 &       2.9$\pm$0.7 &  52 & 17$^{+12}_{-5}$& 34.2/43 & 2$\times$10$^{-3}$\\ 
\end{tabular}
\end{center}
\end{table*}
\begin{table*}
\addtocounter{table}{-1}
\caption{{\it Table \ref{tab:bbfits} continued.}}
\begin{center}
\begin{tabular}{l l l l l l l l l }
\hline \hline
101219B:160-544 &PL& 0.16 f &&       1.71$\pm$0.05 &&&292/186 & \\ 

101219B:160-544 &PL+BB& 0.16 f &      0.20$\pm$0.02 &       1.34$\pm$0.07 &   14 & 3.5$^{+0.4}_{-0.5}$ & 159/184 & 10$^{-25}$\\ 
\hline
101219B:160-256 &PL& 0.16 f &&        1.30$\pm$0.07 &&& 85.1/76 & \\ 
101219B:160-256 &PL+BB& 0.16 f &      0.30$\pm$0.06 &        1.1$\pm$0.1 &  6 & 2$\pm$1& 75.1/74 & 1$\times$10$^{-4}$\\ 
101219B:256-320 &PL& 0.16 f &&       1.62$^{+0.10}_{-0.09}$ &&& 36.5/42 & \\ 
101219B:256-320 &PL+BB& 0.16 f &      0.23$\pm$0.06 &       1.37$^{+0.17}_{-0.18}$ &  9 & 2.5$\pm$1 & 26.8/40 & 2$\times$10$^{-3}$\\ 
101219B:320-544 &PL& 0.16 f &&       2.17$\pm$0.08 &&& 174/89 & \\ 
101219B:320-544 &PL+BB& 0.16 f &      0.18$\pm$0.02 &       1.56$\pm$0.12 &  25 & 4.8$\pm$0.7 & 73.0/87 & 10$^{-17}$\\ 
\hline \hline
120422A:128-192 &PL& 0.027 f && 2.98$\pm$0.06 &&& 213/59 & \\ 
120422A:128-192&PL+BB& 0.027 f& 0.204$\pm$0.009 & 2.37$^{+0.27}_{-0.34}$ & 65 & 1.9$^{+0.2}_{-0.3}$ & 76/57 & 10$^{-13}$\\ 
120422A:128-192 &PL& 0.36$^{+0.07}_{-0.06}$ &&       4.35$^{+0.30}_{-0.27}$ &&& 88/58 & (10$^{-12}$)\\ 
120422A:128-192 &PL+BB& 0.13$^{+0.14}_{-0.11}$ &  0.17$^{+0.03}_{-0.04}$ & 2.91$^{+0.65}_{-0.62}$ &  52 & 2.3$^{+1.8}_{-0.6}$ & 74/56 & 8$\times$10$^{-3}$\\ 
\end{tabular}
\end{center}
\end{table*}

\begin{table*}
\caption{Exponentially cut-off power law (eCPL) and broken power law (BKNPL) fits to those sources requiring further investigation. Column 1 also lists time since BAT trigger covered by the spectrum.}
\label{tab:cutoffpl}
\begin{center}
\begin{tabular}{l l l l l l}
GRB:time & model& $N_{\rm H,int}$& $\Gamma_{1,2}$ & $E_{\rm pk/bk}$ & $\frac{\chi^2}{dof}$\\ 
~~~~~~~~~(s) & & $\times$10$^{22}$ (cm$^{-2}$)&  & (keV) & \\ \hline \hline
060729:128-160 & BKNPL & 0.15 f & 0.8$^{+0.1}_{-0.2}$, 2.7$\pm$0.1 & 1.17$^{+0.05}_{-0.09}$  & 71/71  \\
060729:195-352 & BKNPL & 0.15 f & 2.5$^{+0.1}_{-0.2}$, 3.62$\pm$0.07 & 0.70$^{+0.05}_{-0.06}$&161/116  \\

070419A:112-304 & eCPL & 0.1 f & 2.56$^{+0.07}_{-0.08}$&4.0$^{+0.8}_{-0.4}$ & 166/165\\
070419A:112-304 & eCPL & 0.90$^{+0.08}_{-0.09}$ & 2.4$\pm$0.1 &4.6$^{+3.1}_{-0.9}$ & 164/164\\
070419A:150-304 & eCPL & 0.1 f & 2.68$\pm$0.09&4.8$^{+1.4}_{-0.7}$ & 110/110\\
070419A:150-304 & eCPL & 0.8$\pm$0.1 & 2.4$^{+0.2}_{-0.1}$ & unconstrained  & 105/109\\

080319B:60-1744 & eCPL & 0.12 f& 1.37$^{+0.02}_{-0.03}$ & 3.9$\pm$0.2  &  713/592 \\     
080319B:60-304 & eCPL &0.12 f & 1.60$^{+0.07}_{-0.08}$& 6.6$^{+7.9}_{-1.9}$ & 288/265   \\
080319B:304-704 & eCPL &0.12 f& 1.43$\pm$0.05 & 5.0$^{+0.8}_{-0.6}$  &  413/386  \\
080319B:704-1744 & eCPL &0.12 f& 1.34$\pm$0.02 & 3.8$\pm$0.2  &  694/572  \\

080319B:704-1744 & BKNPL &0.12 f& 1.41$\pm$0.02, 1.87$\pm$0.02 & 1.40$^{+0.09}_{-0.06}$  &  663/571  \\
080319B:704-1744 & BKNPL &0.10$^{+0.02}_{-0.03}$& 1.37$^{+0.05}_{-0.08}$, 1.87$\pm$0.02 & 1.38$^{+0.07}_{-0.05}$ &  661/570  \\

081007:105-175&eCPL&0.7 f& 2.6$\pm$0.1& unconstrained & 45.6/37 \\
100418A:64-160 & BKNPL & 0.2 f& 1.5$^{+0.7}_{-1.1}$, 3.8$\pm$0.2 & 0.6$\pm$0.1  & 40/41\\
101219B:160-544 & eCPL & 0.16 f& 1.71$\pm$0.05 & unconstrained & 292/185 \\
120422A:128-192 & BKNPL & 0.037 f & 0.6$^{+0.5}_{-0.6}$, 3.6$\pm$0.1 & 0.60$\pm$0.04  & 79/57  \\
\end{tabular}
\end{center}
\end{table*}

\begin{figure}
\begin{center}
\caption{Evolution of spectral parameters when fitting a PL+BB model to GRB\,101219B time-sliced spectra (Tables \ref{tab:bbfits},\ref{tab:bbfinallist}). The top panel shows steepening of the underlying power law index, the second panel from the top shows cooling of the fitted BB. The lower two panels show the evolutionary trends of luminosity and radius for the BB with errors taken from the BB normalisation and temperature (redshift uncertainty is not included). Time since BAT trigger is in the observer frame and plotted on a log scale. 
}
\label{bbevol101219b}
\includegraphics[width=8cm, angle=0]{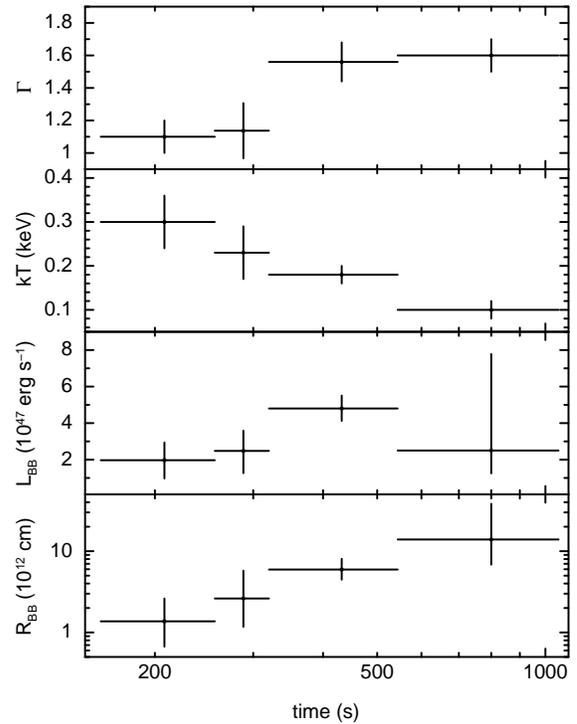}
\end{center}
\end{figure}

\section{Supernova-less GRBs} \label{sec:060614}
An intriguing individual case which fits
all but the first of our sample selection requirements in Section \ref{sec:sample} (the requirement for an associated SN) is GRB\,060614. This source was assigned to the long GRB class (e.g. Xu et al. 2009), while this was somewhat controversial given the short spike$+$long tail prompt emission profile (e.g. Gehrels et al. 2006). It lies at a low redshift, $z = 0.125$ (e.g. Fynbo et al. 2009), yet no optical supernova signature was found to very deep limits (Della Valle et al. 2006; Fynbo et al. 2006; Gal-Yam et al. 2006). Mangano et al. (2007) performed detailed fits to the WT mode data of 060614, always with a variable intrinsic column density. They improved the fit statistic by going from an absorbed PL to either PL+BB, eCPL or Band models (all equally acceptable), and on inclusion of the BAT data during overlap conclude that this was due to the spectral peak energy moving through the XRT band from 97--175\,s after burst. We performed the same fits as in Section \ref{sec:method} on the WT time-averaged (97--466\,s) spectrum of 060614, for consistency and comparison with our GRB-SN sample and in the hope that it may shed further light on the nature of this unusual GRB. We find that a single absorbed PL provides a good fit with no need for further components, when fixing the column densities to $N_{\rm H,Gal} = 1.9 \times 10^{20}$ cm$^{-2}$ (LAB H\,I Survey) and $N_{\rm H,int} = 3 \times 10^{20}$ cm$^{-2}$ (late-time PC mode spectral fit). Allowing the intrinsic column density to vary, the PL model is still preferred over a PL+BB. These results are in fact consistent with those of Mangano et al. (2007): the mean photon arrival time of 250\,s for the time-averaged spectrum means that $E_{\rm pk}$ will no longer be detectable within the XRT energy band, and a single PL will be an adequate fit. 

\begin{table*}
\caption{Measured properties of the final list of BB candidates and possible candidates from the time-averaged spectra as listed in Table \ref{tab:bbfits} (N.B. Fits to GRBs 100316D, 100418A and 120422A are $N_{\rm H,int}$-dependent). For GRB\,101219B we also list the time-sliced spectral results including PC mode.}
\label{tab:bbfinallist}
\begin{center}
\begin{tabular}{l l l l l l l}
GRB &t$_{\rm logmid}$& $T_{\rm BB}$ & $F_{\rm BB}$ & $\frac{F_{\rm BB}}{F_{\rm X}}$&  $L_{\rm BB}$ & $R_{\rm BB}$\\
 & restframe (s) & (keV) & (erg cm$^{-2}$ s$^{-1}$)&(\%) & $\times$10$^{47}$ (erg s$^{-1}$)& (cm) \\ \hline \hline
060218 & 643.9 & 0.181$\pm$0.004 & 8.10$\times$10$^{-10}$ & 13 & 0.024 & 4$\times$10$^{11}$\\
090618 & 114.8 &  0.9$^{+0.3}_{-0.2}$ &  6.530$\times$10$^{-9}$ &10&181&1.5$\times$10$^{12}$ \\
090618 & 129.9 &0.51$^{+0.08}_{-0.05}$ &  2.150$\times$10$^{-9}$ &11&61&3$\times$10$^{12 }$\\
100316D & 307.4 & 0.72$\pm$0.02 &  5.00$\times$10$^{-10}$ &20&0.046&4$\times$10$^{10}$ \\
100316D & & 0.58$^{+0.12}_{-0.06}$ & 1.30$\times$10$^{-10}$ &5&0.012&3$\times$10$^{10}$ \\
100316D & & 0.176$\pm$0.009 & 1.500$\times$10$^{-9}$ &33&0.16&1$\times$10$^{12}$ \\
100316D & & 0.23$^{+0.04}_{-0.03}$& 2.58$\times$10$^{-10}$ &8&0.026& 3$\times$10$^{11}$ \\
101219B &  202.4 & 0.30$\pm$0.06 &6.1$\times$10$^{-11}$ &6&1.97 &9$\times$10$^{11}$\\
101219B & 286.2 & 0.23$\pm$0.06 &6.9$\times$10$^{-11}$ &9&2.48 & 2$\times$10$^{12}$ \\
101219B & 417.2 & 0.18$\pm$0.02 &1.13$\times$10$^{-10}$ &25&4.80 &4$\times$10$^{12}$ \\
101219B & 757.9 & 0.10$\pm$0.02 & 2.6$\times$10$^{-11}$ &34&2.50 &  7$\times$10$^{12}$ \\ \hline
060729 & 169.8&0.21$\pm$0.01 & 6.31$\times$10$^{-10}$ &37& 22&9$\times$10$^{12}$\\
081007 & 88.6& 0.31$\pm$0.04& 2.10$\times$10$^{-10}$ &32& 6 & 2$\times$10$^{12}$\\
100418A & 62.3 & 0.23$\pm$0.01&3.00$\times$10$^{-10}$ &57&16&7$\times$10$^{12}$\\
100418A & & 0.22$\pm$0.04&3.15$\times$10$^{-10}$ &52&17&8$\times$10$^{12}$\\
120422A & 122.5& 0.204$\pm$0.009& 3.98$\times$10$^{-10}$&65& 1.9&3$\times$10$^{12}$\\
120422A & &0.17$^{+0.03}_{-0.04}$& 4.49$\times$10$^{-10}$ &52&2.3 &5$\times$10$^{12}$\\
\end{tabular} 
\end{center}
\end{table*}

\section{Discussion} \label{sec:discussion}
We report a new thermal X-ray component detection for GRB\,101219B associated with SN2010ma, which we believe to be compelling. We also recover the BB components reported previously in GRBs 060218, 090618 and 100316D with consistent parameters. For our remaining 5 BB candidate sources from Table \ref{tab:bbfits} (F-test statistic $\le 10^{-3}$) we then omitted times of major flaring in the light curves, time-sliced according to light curve segment, allowed $N_{\rm H,int}$ to vary when the error from the late-time measurement was very large and fit eCPL and BKNPL models. We performed joint BAT-XRT fits to all our sample where possible to locate the spectral peak energy and thereby better diagnose a spectral break in the early XRT data.

We found we could reasonably explain the spectral curvature with one of the eCPL or BKNPL models for GRB\,070419A and the early time-slice of GRB\,060729. We conclude that four sources may show thermal X-ray emission within their early X-ray spectra, but cannot rule out other explanations for this curvature. The reasons for this are most often because of large uncertainties in absorbing column and degeneracy between this and the underlying spectrum, or unconfirmed spectral breaks in the spectra. 

\noindent
{\bf The role of X-ray absorbing columns}\\The absorbing column along the line-of-sight towards a GRB can be substantial, and reduces the observed X-ray flux in the region we are interested in for BB detection. It is therefore likely that details in the soft X-ray spectrum will be missed if the absorbing column is sufficiently high. The column density limit above which this happens is investigated in Paper II and must depend on power law index and source redshift. We note that the Galactic column towards our GRB-SNe has a mean of 4$\times$10$^{20}$ cm$^{-2}$, which is insignificant when compared with the mean intrinsic measured column of $\sim$4$\times$10$^{21}$ cm$^{-2}$ and so should have no effect. Signal to noise in many of our spectra is not good enough to keep $N_{\rm H,int}$ free and add a BB, hence in the first fits in this work we kept $N_{\rm H,int}$ fixed at the late-time value (Table \ref{tab:knownproperties}). This is clearly not ideal, and we had to
identify the cases with very uncertain late-time absorbing column densities, $N_{\rm H,int}$, and allow for this. The BB in 100418A must be deemed uncertain due to the errors on $N_{\rm H,int}$, while 120422A is consistent with having very little or no intrinsic absorption. For the remaining sources, we note that 060218, 081007 and possibly 100316D all have greater intrinsic columns than the mean of the sample and, while the BB is clear in 060218 and 100316D, it is likely that the high column ($N_{\rm H,int}~=~$(0.7$\pm$0.2) $\times 10^{22}$ cm$^{-2}$) is affecting our ability to either securely identify or rule out a BB component in 081007. 

\noindent
{\bf Spectral evolution}\\
Spectral evolution at early times in the X-ray production of GRBs is common, where the power law slope of the spectrum is seen to steepen through the steep decay phase in the light curve (e.g. Zhang, Liang \& Zhang 2007). We can estimate the effect of spectral evolution on our sources by examination of their hardness ratios as created in the light curves section of the UKSSDC Swift XRT GRB Repository$^1$, and in some cases we split the spectra according to the different light curve segments (different decay slopes). 
Hardening during flares is a particular problem and we have done our best to eliminate flaring episodes from our spectra. The time-sliced analysis of GRB\,101219B shows a softening of the power law, but this did not render the BB undetectable in the time-averaged data so we expect this effect will not be a significant issue.   

The presence of the spectral peak energy in the X-ray band during our observations would invalidate our assumption of a single underlying absorbed power law continuum, requiring a spectral break or added curvature for the case of a long-duration exposure in which the break energy is seen to move. We know that $E_{\rm pk}$ is present in the X-ray spectrum we use here for GRB\,060218 yet we can still conclusively recover the BB component (though it is evident from the poor fit statistic that the continuum model remains incorrect, likely due to significant spectral evolution). This was the most prominent thermal component ever detected, comprising up to half of the observed X-ray flux. The exponentially cut-off power law fits to the spectra (Table \ref{tab:cutoffpl}), together with information from either prompt gamma-ray fits or our own BAT-XRT fits (Table \ref{tab:epk}) have allowed us to reasonably assess the need for a spectral break in each spectrum. Indeed, we find that a break is likely to be present in the early X-ray data of 070419A, 060729 pre-flare and 080319B (in the latter we note that a two-jet model suggested in Racusin et al. 2008 rather implies a superposition of power laws, but other interpretations, namely that of Bloom et al. 2009, require only one jet).  

\noindent
{\bf Instrumental effects}\\
We note that the quasar 3C\,273 is, like some of our GRB-SN, well described by an absorbed power law plus blackbody and may be used to assess the effect of calibration uncertainties on our sample. This source is a regular {\it Swift} XRT calibration target, and was observed in 2005 simultaneously with {\it XMM-Newton} and {\it RXTE} for cross-calibration purposes. The BB in 3C\,273 has a temperature of $kT=0.1$\,keV and contributes 4\% of the XRT X-ray flux. The spectral fits for XRT WT mode agree to within the expected uncertainties with the {\it XMM} EPIC and {\it RXTE} spectra (we note that for high statistical spectra a systematic error of $\le$3\% was found to be sufficient, see e.g. SWIFT-XRT-CALDB-09\_v11 and subsequent calibration release notes). This gives us confidence that the BB components required in the GRB-SN sample are real spectral features, and cannot be an instrumental effect.

\subsection{Comparison with GRB, SN and host galaxy properties} \label{sec:broadband}
Although we have small number statistics here with only 11 sources in our sample, we can examine the GRB, SN and host galaxy properties and look for commonalities between the source and BB properties. We split our GRBs into three `detection categories': detected BB, possible BB and no BB, as they are set out in Section \ref{sec:results}. No correlations arise between any of the properties we looked at: $z$, $T_{90}$, $S_{\rm BAT 15-150\,keV}$, $E_{\rm iso}$, t$_{\rm obs,start}$, t$_{\rm mid,rest}$, $\alpha_{\rm X}$, $N_{\rm H,Gal}$, $N_{\rm H,int}$, $\Gamma$, $kT_{\rm BB}$, $F_{\rm BB}$/$F_{\rm X}$, $L_{\rm BB}$, $R_{\rm BB}$, F-statistic, $M_{\rm V,SN-peak}$, SN detection technique (spectroscopic or photometric), $M_{\rm B,host}$ when host is detected and BB required (yes, possibly or no).

There is, unsurprisingly, a general trend with redshift: the two most distant sample sources at $z\sim1$ do not have detectable BB signatures, while BBs in the two very nearby sources at $z<0.1$ are clearly detected. A 0.2\,keV BB will peak at $\sim$0.1\,keV for our $z\sim1$ GRBs, leaving just a portion of the BB tail detectable in the XRT range. This may explain the lack of BB detection in GRBs 070419A and 080319B, compounded by high absorbing column density in 070419A. Intrinsic column is likely to play an important role in BB detectability as discussed above and in Paper II. 
Power law photon index also shows a trend, with values of $\Gamma \sim 1.4-1.6$ in the detected category while we find $\Gamma \sim 2.1-2.9$ in the possible BB category and something in between, $\Gamma \sim 1.8-2.2$, for the GRBs with no detected BB. Since the early power law index often evolves from hard to soft in GRBs (e.g. Zhang et al. 2007), it may be that the possible BBs are detected at later stages in the development of the prompt emission. The steeper power law indices make determination of the slopes more difficult and typical errors on $\Gamma$ are $\pm >0.2$ compared with $\pm \le 0.1$ for the detected BB category sources. The temporal power law index, however, has no bearing on the detection of a BB, with various light curve types sampled in each category (Figure \ref{lcs}). Considering the prompt gamma-ray properties of these sources (as measured with {\it Swift} BAT for consistency across all GRBs, Sakamoto et al. 2011), while the first two BBs were discovered in subenergetic, long-duration GRBs, the candidates proposed since then have extended into the typical cosmological GRB range. GRB\,090618 is still the most energetic at $E_{\rm iso} \sim 8 \times 10^{52}$ erg, and the span of $E_{\rm iso}$ in the rest of our candidates is 10$^{49}$-10$^{51}$ erg. This strongly implies that thermal X-ray emission (or significant spectral curvature) can arise in any GRB, and its detection is dependent upon properties other than the prompt gamma-ray emission. 

We notice a distinctly lower BB luminosity and radius for the two low $E_{\rm iso}$ GRBs 060218 and 100316D with detected BBs when compared with the GRBs with possible BBs, while GRB 101219B appears to lie in between the two categories. GRB 090618 does not follow the trend, with a very high BB luminosity, unsurprising given its very high afterglow flux, but a relatively low inferred radius. The one BB parameter that correlates well with detection category is the fractional unabsorbed 0.3-10 keV X-ray flux contributed by the BB. This lies between 8 and 14\% for the unambiguous cases, and between 32\% and 65\% in all possible detections. Interestingly, the restframe temperature of the BB remains similar for all detections, at around 0.2\,keV, with the exception of 090618 in which we see a somewhat higher $kT$, decreasing from 0.9--0.5\,keV.

 For a number of GRB-SNe the SN properties have been measured
(Hjorth \& Bloom 2011; Cano et al. 2011b; Bufano et al. 2012; Sparre et
al. 2011; Cobb et al. 2010; de Ugarte Postigo et al. 2011, in preparation; Richardson 2009; Ferrero et al. 2006; Melandri et al. 2012; Schulze et al. in preparation), but beyond SN peak
absolute magnitude the
overlap with this sample is small since we also include photometric-only SN
detections for which detailed ejecta information is not available. 
The GRBs with spectroscopically-determined supernovae are 060218, 081007, 091127, 100316D, 100418A, 101219B and 120422A, spanning all 3 categories of BB detection. SN peak absolute magnitudes are not available for 070419A, 081007 and 101219B. Host galaxy magnitude, which may be used as an approximate tracer for metallicity, is available for around half the sample: 060218, 060729, 080319B, 090618, 091127 and 100316D hampering comparison with other properties. We can only say that 2 out of 3 GRBs with a clear BB have fainter hosts than the 3 possible/no-BB GRBs, the bright host of 060218 being the exception.
 
\subsection{Origins of the thermal X-ray component} \label{modelling}
The evolution of spectral parameters in our fits to GRB\,101219B (Figure \ref{bbevol101219b}) may be compared with similar analyses performed for GRBs 060218 and 100316D (Figure 4 of Starling et al. 2011; Figure 4 of Olivares et al. 2012; see also Kaneko et al. 2007) and GRB\,090618 (Figure 8 of Page et al. 2011). While the passage of the spectral peak energy through the X-ray band in the very soft GRBs 060218 and 1001316D complicates the comparison, GRB\,101219B shows a more similar behaviour to 090618 for most spectral parameters. The exception is BB luminosity, which steadily decreased in 090618 over 125--275\,s since trigger. We find a general increase in 101219B from 160--544\,s, followed by a decrease (though this decrease is only seen in the PC spectra which we stress can only be used as a guide to the BB behaviour). The increase does coincide with a flattening of the light curve which may be interpreted as a short plateau, flaring episode or other transition phase between the $\alpha \sim 1$ and $\alpha \sim 2$ decays which may complicate the interpretation of our results. The luminosity variation has, however, little effect on the radius determination, which is almost exactly the same value and trend as is seen in 090618, increasing from 9$\times$10$^{11}$\,cm to 7$\times$10$^{12}$\,cm in at most $\sim$900\,s, (mean of 556\,s) implying an expansion velocity of 0.4$^{+0.3}_{-0.2}$c. Firstly, the radii that result are large compared with the 10$^{11}$\,cm estimated for Wolf-Rayet stars (e.g. Cappa et al. 2004), a likely progenitor type for long GRBs, but the existence of a thick wind could increase the radius probed in this way, as was suggested by some authors for the previous thermal X-ray components within the shock breakout model (but see e.g. Li 2007). Secondly, the inferred expansion velocity is very high, at least 20\% of the speed of light, and together these considerations cast some doubt on the origin of this thermal emission in the shock breakout of GRB\,101219B. We reiterate that these radii are illustrative only, since they assume both a single temperature BB and thermal equilibrium: at the high luminosities recovered here it seems likely a shock breakout would be highly relativistic, where thermal equilibrium does not hold and indeed the predicted temperatures put the observable signature at energies above the soft X-ray band (Nakar \& Sari 2012, before consideration of the effects of a wind). 

An alternative origin for thermal emission may be in a cocoon surrounding the jet. A number of our BB candidates are discovered during the steep decay phase of the X-ray light curve (Figure \ref{lcs}). Pe'er, M\'esz\'aros \& Rees (2006) showed that steep X-ray light curve decays ($\alpha > 2$) can be produced by a relativistically expanding hot plasma cocoon associated with the jet. They demonstrate that some time after the prompt emission, typically a few hundred seconds, the bulk of the cocoon emission would fall in the X-ray band. 
The original spectrum is assumed to be a power law with power law index $p=2$,
as is expected, e.g., from synchrotron radiation. However, during the
expansion phase of the cocoon the photons lose energy to the expanding
plasma, and the spectrum is modified, resulting in a low energy thermal
component combined with a power law over the 0.3--10\,keV range.
Origin in such a cocoon would allow measured BB radii to be larger than is expected for shock breakout, $\ge$10$^{12}$\,cm, in line with many of our measurements. The high luminosities of the BB candidates would also be more easily explained in this model compared with the shock breakout model. 
The eventual predicted temporal X-ray decay is steep in this model, $\alpha \ge 2$, which applies to some of our sample GRBs: 060729, 081007, 090618, 100418A, 101219B and 120422A. Interestingly, almost flat early light curves are seen in GRBs 060218 and 100316D, i.e. in the most convincing cases thus far for the shock breakout model. Additionally, very steep decays do not occur in the data in-hand for 070419A, 080319B or 091127, all GRBs where a BB could not be detected. 

\begin{figure}
\begin{center}
\caption{Here we show models for emission from a relativistically expanding hot plasma cocoon associated with the jet, described in Pe'er et al. (2006, see Section \ref{modelling}). These demonstrate that this cocoon model can produce the general shape seen in the spectrum (upper panel) and light curve (lower panel) of some of our GRB-SN thermal X-ray emission candidates. These are not fits to the data, but parameters are optimised for best representation of GRB\,101219B during T$_{\rm 0}+100-250$\,s. In each panel the three curves represent possible choices for model parameters. Cyan: optical depth $\tau=30$, initial expansion radius $r_{\rm initial} = 5\times 10^{12}$\,cm, total cocoon energy $E_{\rm c} = 5 \times 10^{49}$ erg, $\Gamma = 4.0$; black: $\tau = 30$, $r_{\rm initial} = 6 \times 10^{12}$\,cm, $E_{\rm c} = 4.2 \times 10^{49}$ erg, $\Gamma = 4.5$; red: $\tau = 50$, $r_{\rm initial} = 6 \times 10^{12}$\,cm, $E_{\rm c} = 4.2 \times 10^{49}$ erg, $\Gamma = 6.5$.}
\label{cocoonmodel}
\includegraphics[width=7cm, angle=0]{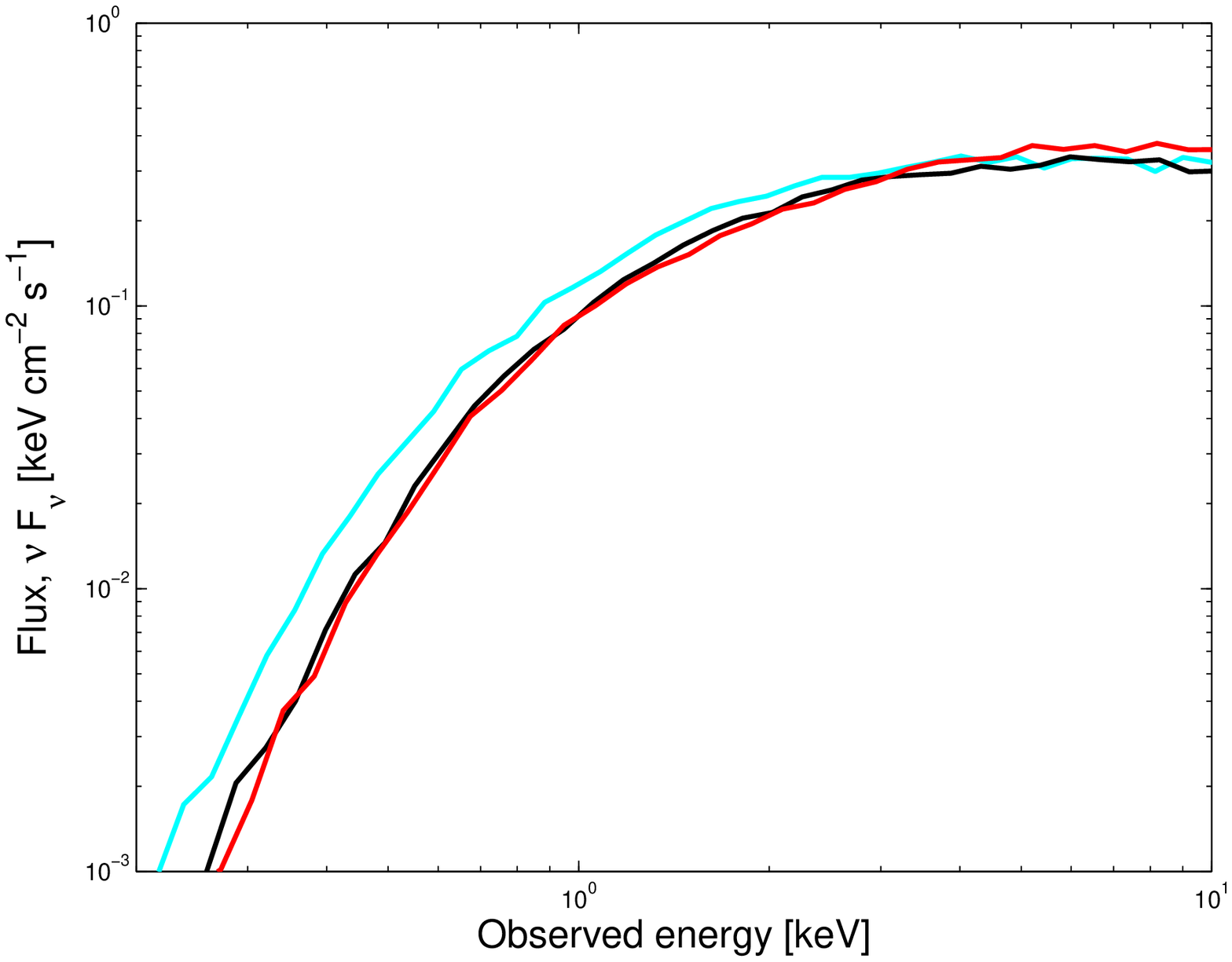}
\includegraphics[width=7cm, angle=0]{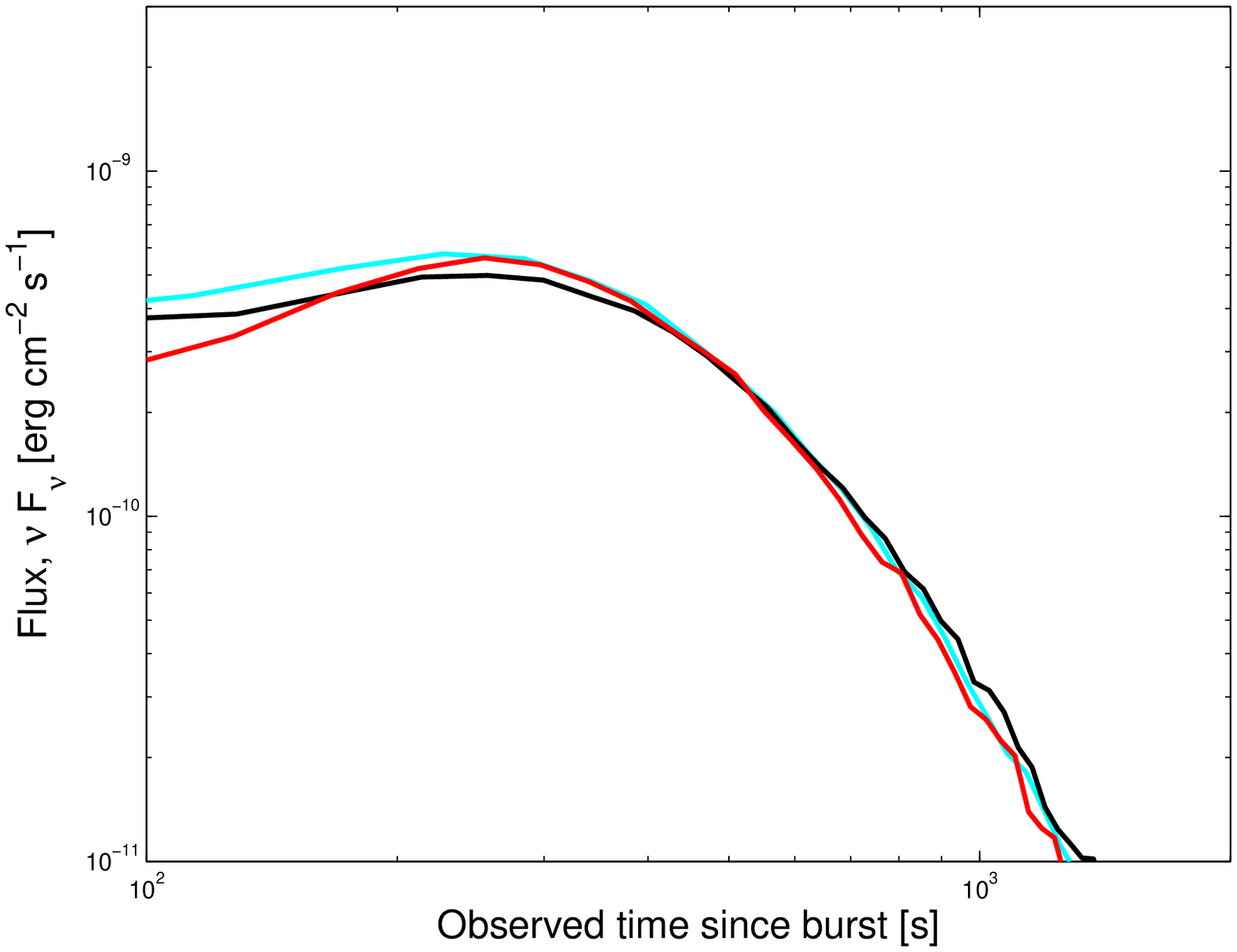}
\end{center}
\end{figure}

Comparing the cocoon model of Pe'er et al. (2006) to our data for GRB\,101219B we find that for reasonable model parameters both the overall spectral shape and the light curve decay 
can be reproduced. We have restricted the modelling to the time frame 100--250\,s since trigger in this first approximate analysis, simply to see whether or not the general trends can be recovered. In Figure \ref{cocoonmodel} we show some of these models, which stem from photons injected deep in the flow having an initial photon energy distribution of a power law with index $p=2$ above 10$^{−3}$ m$_e$ c$^2$. The resulting spectrum comprises low energy thermal spectra with modified power laws at higher energies, and the steep flux decay is produced, when the initial optical depth of the cocoon, $\tau$, is 30--50. The Lorentz factor of the cocoon itself is found to be $\Gamma = 4-7$ with initial cocoon radius $\sim 5 \times 10^{12}$ cm$^{-2}$. Total cocoon energy we have adopted, adjusted to match the light curve and energy of the early X-ray data, is of order a few $\times 10^{49}$ erg ($E_{\rm iso}$ as measured from prompt gamma-ray data is $\sim 5 \times 10^{51}$\,erg).

\section{Conclusions} \label{sec:concl}
We have performed a systematic search for blackbody thermal components in the early {\it Swift} X-ray spectra of 11 GRBs with associated supernovae. This has resulted in the detection of 4 compelling cases where, in addition to an absorbed power law, a BB-like component is required. We also note 4 possible BB candidates and their caveats and 3 GRB-SN without detectable thermal emission. Amongst the detections, we report for the first time details for GRB\,101219B in which we see a cooling BB of restframe temperature $\sim$0.2 keV, luminosity $\sim$3$\times$10$^{47}$ erg s$^{-1}$ and inferred radius $\sim$4$\times$10$^{12}$ cm expanding at 0.2c or greater. Its general properties appear to be intermediate between the low luminosity GRBs 060218 and 100316D and the classical GRB\,090618, all of which have previously reported thermal X-ray emission that we also recover here. 

We cannot identify any strong correlations of BB properties with the available prompt emission, SN or host properties. We note some trends between BB `category' and redshift, power law spectral index and percentage contribution of the BB to the 0.3--10\,keV flux; these likely do not refer to the intrinsic attributes of the GRBs, but rather to the detachability of the thermal component. All the possible BB candidates among our sample have large BB luminosities ($>10^{47}$ erg cm$^{-2}$) and inferred radii ($>10^{12}$ cm). These luminosities are difficult to reconcile with shock breakout models since they imply highly relativistic outflows that would emit largely at energies above the {\it Swift } XRT bandpass. Our BB candidates occur during the steep decay phase of the X-ray light curve. We discuss a model for this emission arising in a relativistically expanding hot plasma cocoon associated with the jet. This model is able to generally reproduce the extreme BB parameters we retrieve from our fits, and is consistent with a steep temporal decay. Only GRBs 060218 and 100316D do not show a temporal decay of $\alpha \ge 2$ during the first $\sim$1000\,s, where the BB luminosities and radii are significantly lower than in the rest of our GRB-SN sample, and the shock breakout model has been argued by a number of authors to be valid (Section \ref{sec:intro}). Indeed, these two low luminosity GRBs may be the product of quite different physics early on from the classical energetic GRBs that populate most of our sample, with GRB\,101219B lying somewhere in the transition.

\section{Acknowledgments}
We thank Claudio Pagani, Antonio de Ugarte Postigo and Klaas Wiersema for useful discussions. RLCS is supported by a Royal Society Dorothy Hodgkin Fellowship. KLP, APB and JPO acknowledge financial support from the UK Space Agency for the \emph{Swift} project. This work made use of data
supplied by the UK {\it Swift} Science Data Centre at the University of Leicester.

\bsp

\label{lastpage}


\begin{thebibliography}{References}
\bibitem[\protect\citeauthoryear{Balberg \& Loeb}{2011}]{Balberg}
Balberg S., Loeb A., 2011, MNRAS, 414, 1715
\bibitem[\protect\citeauthoryear{Berger et al.}{2008}]{Berger}
Berger E., Fox D.B., Cucchiara A., Cenko, S.B., 2008, GCN 8335
\bibitem[\protect\citeauthoryear{Berger et al.}{2011}]{Berger2}
Berger E. et al., 2011, ApJ, 743, 204
\bibitem[\protect\citeauthoryear{Bissaldi et al.}{2008}]{Bissaldi}
Bissaldi E., McBreen S., Connaughton V., 2008, GCN Circ. 8369
\bibitem[\protect\citeauthoryear{Bloom et al.}{2009}]{Bloom}
Bloom J.S. et al., 2009, ApJ, 691, 723
\bibitem[\protect\citeauthoryear{Bo\"er et al.}{2006}]{Boer}
Bo\"er M., Atteia J.L., Damerdji Y., Gendre B., Klotz A., Stratta G., 2006, ApJ, 638, L71
\bibitem[\protect\citeauthoryear{Bromberg, Nakar \& Piran}{2011}]{Bromberg}
Bromberg O., Nakar E., Piran T, 2011, ApJ (Letters), 739, L55
\bibitem[\protect\citeauthoryear{Bromberg et al.}{2012}]{Bromberg2}
Bromberg O., Nakar E., Piran T., Sari R., 2012, ApJ, 749, 110
\bibitem[\protect\citeauthoryear{Bufano et al.}{2012}]{Bufano}
Bufano F. et al., 2012, ApJ, 753, 67 
\bibitem[\protect\citeauthoryear{Burrows et al.}{2005}]{Burrows}
Burrows D.N. et al., 2005, Space Sci. Rev., 120, 165
\bibitem[\protect\citeauthoryear{Butler \& Kocevski}{2007}]{ButlerKocevski}
Butler N., Kocevski D., 2007, ApJ, 663, 407
\bibitem[\protect\citeauthoryear{Campana et al.}{2006}]{Campana}
Campana S. et al., 2006, Nature, 442, 1008 
\bibitem[\protect\citeauthoryear{Campana et al.}{2007}]{Campana07}
Campana S. et al., 2007, ApJ, 654, L17
\bibitem[\protect\citeauthoryear{Campana et al.}{2011a}]{Campana2}
Campana S. et al., 2011a, Nature, 480, 69
\bibitem[\protect\citeauthoryear{Campana et al.}{2011b}]{Campana3}
Campana S., D'Avanzo P., Lazzati D., Covino S., Tagliaferri G., Panagia N., 2011b, MNRAS, 418, 1511
\bibitem[\protect\citeauthoryear{Cano et al.}{2011a}]{Cano1}
Cano Z. et al., 2011a, MNRAS, 413, 669
\bibitem[\protect\citeauthoryear{Cano et al.}{2011b}]{Cano2}
Cano Z. et al., 2011b, ApJ, 740, 41
\bibitem[\protect\citeauthoryear{Cappa et al.}{2004}]{Cappa}
Cappa C., Goss W. M., van der Hucht K.A., 2004, AJ, 127, 2885
\bibitem[\protect\citeauthoryear{Cash}{1979}]{Cash}
Cash W., 1979, ApJ, 228, 939
\bibitem[\protect\citeauthoryear{Chevalier \& Fransson}{2008}]{Chevalier}
Chevalier R.A., Fransson C., 2008, ApJ (Letters), 683, L135
\bibitem[\protect\citeauthoryear{Cobb et al.}{2010}]{Cobb}
Cobb B.E., Bloom J.S., Perley D.A., Morgan A.N., Cenko S.B., Filippenko A.V., 2010, ApJ (Letters), 718, L150
\bibitem[\protect\citeauthoryear{Colgate}{1974}]{Colgate}
Colgate S.A., 1974, ApJ, 187, 333
\bibitem[\protect\citeauthoryear{Couch et al.}{2011}]{Couch}
Couch S.M., Pooley D., Wheeler C.J., Milosavljevi\'c M., 2011, ApJ, 727, 104 
\bibitem[\protect\citeauthoryear{Cucchiara et al.}{2009}]{Cucchiara}
Cucchiara A., Fox D., Levan A., Tanvir N., 2009, GCN Circ. 10202
\bibitem[\protect\citeauthoryear{Della Valle et al.}{2006}]{DellaValleNature}
Della Valle M. et al., 2006, Nature, 444, 1050
\bibitem[\protect\citeauthoryear{Della Valle et al.}{2006}]{DellaValle}
Della Valle M. et al., 2006, ApJ (Letters), 642, L103
\bibitem[\protect\citeauthoryear{Della Valle et al.}{2008}]{DellaValle2}
Della Valle M. et al., 2008, CBET 1602
\bibitem[\protect\citeauthoryear{de Ugarte Postigo et al.}{2011a}]{deugarte}
de Ugarte Postigo A., Th\"{o}ne C.C., Goldoni P., Fynbo J.P.U., 2011a, Astronomische Nachrichten, 332, 297
\bibitem[\protect\citeauthoryear{de Ugarte Postigo et al.}{2011b}]{deugarte2}
de Ugarte Postigo A., Goldoni P., Milvang-Jensen B., Malesani D., Sparre
M., Fynbo J.P.U., Leloudas G., Covino S., Flores H., D'Elia V., Levan A.,
2011b, GCN Circ. 11579
\bibitem[\protect\citeauthoryear{de Ugarte Postigo et al.}{2012}]{deugarte3}
de Ugarte Postigo A., Thoene C.C., Gorosabel J., 2012, GCN Circ. 12802
\bibitem[\protect\citeauthoryear{Evans et al.}{2007}]{Evans}
Evans P.A. et al., 2007, A\&A, 469, 379
\bibitem[\protect\citeauthoryear{Evans et al.}{2009}]{Evans2}
Evans P.A. et al., 2009, MNRAS, 397, 1177
\bibitem[\protect\citeauthoryear{Ferrero et al.}{2006}]{Ferrero}
Ferrero P. et al., 2006, A\&A, 457, 857
\bibitem[\protect\citeauthoryear{Fynbo et al.}{2006}]{Fynbo}
Fynbo J.P.U. et al., 2006, Nature, 444, 1047
\bibitem[\protect\citeauthoryear{Fynbo et al.}{2009}]{Fynboz}
Fynbo J.P.U. et al., 2009, ApJS, 185, 526
\bibitem[\protect\citeauthoryear{Gal-Yam et al.}{2006}]{Gal-Yam}
Gal-Yam A. et al., 2006, Nature, 444, 1053
\bibitem[\protect\citeauthoryear{Gehrels et al.}{2004}]{Gehrels}
Gehrels N. et al., 2004, ApJ, 611, 1005
\bibitem[\protect\citeauthoryear{Gendre et al.}{2007}]{Gendre}
Gendre B., Galli A., Corsi A., Klotz A., Piro L., Stratta G.; Bo\"er M., Damerdji Y., 2007, A\&A, 462, 565
\bibitem[\protect\citeauthoryear{Gezari et al.}{2008}]{Gezari}
Gezari S. et al., 2008, ApJ (Letters), 683, L131
\bibitem[\protect\citeauthoryear{Gezari et al.}{2010}]{Gezari2}
Gezari S. et al., 2010, ApJ (Letters), 720, L77
\bibitem[\protect\citeauthoryear{Ghisellini, Ghirlanda \& Tavecchio}{2007}]{Ghisellini}
Ghisellini G., Ghirlanda G., Tavecchio F., 2007, MNRAS, 382, L77
\bibitem[\protect\citeauthoryear{Hill et al.}{2007}]{Hill}
Hill J. et al., 2007, GCN Circ. 6486
\bibitem[\protect\citeauthoryear{Hjorth \& Bloom}{2011}]{HjorthBloom}
Hjorth J., Bloom J., 2011, in $``$Gamma-Ray Bursts$''$, Eds. C. Kouveliotou, R.A.M.J. Wijers, S.E. Woosley, Cambridge University Press, 2011, Chapter 9
\bibitem[\protect\citeauthoryear{Hurkett et al.}{2008}]{Hurkett}
Hurkett C.P. et al., 2008, ApJ, 679, 587
\bibitem[\protect\citeauthoryear{Jakobsson et al.}{2012}]{Jakobsson}
Jakobsson P. et al., 2012, ApJ, 752, 62
\bibitem[\protect\citeauthoryear{Kalberla et al.}{2005}]{Kalberla}
Kalberla P.M.W., Burton W.B., Hartmann Dap, Arnal E.M., Bajaja E., Morras R., P\"oppel W.G.L., 2005, A\&A, 440, 775
\bibitem[\protect\citeauthoryear{Kaneko et al.}{2007}]{Kaneko}
Kaneko Y. et al., 2007, ApJ, 654, 385
\bibitem[\protect\citeauthoryear{Li}{2007}]{Li}
Li L.-X., 2007, MNRAS, 375, 240
\bibitem[\protect\citeauthoryear{Markwardt et al.}{2008}]{Markwardt}
Markwardt C.M. et al., 2008, GCN Circ. 8338
\bibitem[\protect\citeauthoryear{Mazzali et al.}{2006}]{Mazzali}
Mazzali P.A. et al., 2006, Nature, 442, 1018
\bibitem[\protect\citeauthoryear{Melandri et al.}{2012}]{Melandri}
Melandri A. et al., 2012, A\&A, submitted, arXiv:1206.5532
\bibitem[\protect\citeauthoryear{Mirabal \& Halpern}{2006}]{Mirabal}
Mirabal N., Halpern J.P., 2006, GCN Circ. 4792
\bibitem[\protect\citeauthoryear{Modjaz et al.}{2009}]{Modjaz}
Modjaz M. et al., 2009, ApJ, 702, 226
\bibitem[\protect\citeauthoryear{Nakar \& Sari}{2010}]{Nakar}
Nakar E., Sari R., 2010, ApJ, 725, 904
\bibitem[\protect\citeauthoryear{Nakar \& Sari}{2012}]{Nakar2}
Nakar E., Sari R., 2012, ApJ, 747, 88
\bibitem[\protect\citeauthoryear{Ofek et al.}{2010}]{Ofek}
Ofek E.O. et al., 2010, ApJ, 724, 1396
\bibitem[\protect\citeauthoryear{Olivares et al.}{2012}]{Olivares}
Olivares E. F. et al., 2012, A\&A, 539, 76
\bibitem[\protect\citeauthoryear{Pe'er, M\'esz\'aros \& Rees}{2006}]{Peer}
Pe'er A., M\'esz\'aros P., Rees M.J., 2006, ApJ, 652, 482
\bibitem[\protect\citeauthoryear{Pian et al.}{2006}]{Pian}
Pian E. et al., 2006, Nature, 442, 1011
\bibitem[\protect\citeauthoryear{Piro et al.}{2005}]{Piro}
Piro L. et al., 2005, ApJ, 623, 314
\bibitem[\protect\citeauthoryear{Sakamoto et al.}{2011}]{Sakamoto}
Sakamoto T. et al., 2011, ApJS, 195, 2
\bibitem[\protect\citeauthoryear{Schawinski et al.}{2008}]{Schawinski}
Schawinski K. et al., 2008, Science, 321, 223
\bibitem[\protect\citeauthoryear{Soderberg et al.}{2008}]{Soderberg}
Soderberg A.M. et al., 2008, Nature, 453, 469
\bibitem[\protect\citeauthoryear{Soderberg, Berger \& Fox}{2008}]{SoderbergBergerFox}
Soderberg A., Berger E., Fox D., 2008, GCN Circ. 8661
\bibitem[\protect\citeauthoryear{Soderberg et al.}{2010}]{Soderberg2}
Soderberg A.M. et al., 2010, Nature, 463, 513
\bibitem[\protect\citeauthoryear{Sollerman et al.}{2006}]{Sollerman}
Sollerman J. et al. 2006, A\&A, 454, 503
\bibitem[\protect\citeauthoryear{Sollerman et al.}{2007}]{Sollerman2}
Sollerman J. et al., 2007, A\&A, 466, 839
\bibitem[\protect\citeauthoryear{Sparre et al.}{2011}]{Sparre}
Sparre M. et al., 2011, ApJ (Letters), 735, L24
\bibitem[\protect\citeauthoryear{Sparre \& Starling}{2012}]{Sparre2}
Sparre M., Starling R.L.C., 2012, MNRAS in press, arXiv:1207.1447
\bibitem[\protect\citeauthoryear{Starling et al.}{2011}]{Starling}
Starling R.L.C. et al., 2011, MNRAS, 411, 2792
\bibitem[\protect\citeauthoryear{Suzuki \& Shigeyama}{2010}]{Suzuki}
Suzuki A., Shigeyama T., 2010, ApJ, 719, 881
\bibitem[\protect\citeauthoryear{Tanvir et al.}{2012}]{Tanvir}
Tanvir N.R., Levan A.J., Cucchiara A., Fox D.B., 2012, GCN Circ. 13251
\bibitem[\protect\citeauthoryear{Thoene et al.}{2009}]{Thoene}
Thoene C.C. et al., 2009, GCN Circ. 10233
\bibitem[\protect\citeauthoryear{Th\"one et al.}{2011}]{Thoene2}
Th\"one C.C. et al., 2011, Nature, 480, 72
\bibitem[\protect\citeauthoryear{Troja et al.}{2012}]{Troja}
Troja E. et al., 2012, ApJ submitted, arXiv:1201.4181
\bibitem[\protect\citeauthoryear{Ukwatta et al.}{2010}]{Ukwatta}
Ukwatta T.N. et al., 2010, GCN Circ. 10615 
\bibitem[\protect\citeauthoryear{Vergani et al.}{2011}]{Vergani}
Vergani S.D. et al., 2011, A\&A, 535, 127
\bibitem[\protect\citeauthoryear{Verner et al.}{1996}]{Verner}
Verner D.A., Ferland G.J., Korista K.T., Yakovlev D.G., 1996, ApJ, 465, 487 
\bibitem[\protect\citeauthoryear{Watson et al.}{2007}]{Watson}
Watson D., Hjorth J., Fynbo J.P.U., Jakobsson P., Foley S., Sollerman J., Wijers R.A.M.J., 2007, ApJ, 660, L101
\bibitem[\protect\citeauthoryear{Waxman et al.}{2007}]{Waxman}
Waxman E., M\'esz\'aros P., Campana S., 2007, ApJ, 667, 351
\bibitem[\protect\citeauthoryear{Wiersema et al.}{2012}]{Wiersema}
Wiersema K. et al., 2012, GCN Circ. 13276
\bibitem[\protect\citeauthoryear{Wilms, Allen \& McCray}{2000}]{Wilms}
Wilms J., Allen A., McCray R., 2000, ApJ, 542, 914
\bibitem[\protect\citeauthoryear{Xu et al.}{2009}]{Xu}
Xu D. et al., 2009, ApJ, 696, 971
\bibitem[\protect\citeauthoryear{Zhang, Liang \& Zhang}{2007}]{ZhangLiang}
Zhang B.-B., Liang E.-W., Zhang B., 2007, ApJ, 666, 1002
\bibitem[\protect\citeauthoryear{Zhang et al.}{2012}]{Zhang}
Zhang B.-B., Fan Y.-Z., Shen R.-F., Xu D., Zhang F.-W., Wei D.-M., Burrows D.N., Zhang B., 2012, ApJ in press, arXiv:1206.0298
\end{thebibliography}
\end{document}